\documentclass[aps,preprintnumbers,amsmath,amssymb,nofootinbib,eqsecnum,preprintnumbers]{revtex4}

\usepackage{graphicx} 
\usepackage[a4paper,margin=1.25in]{geometry}
\usepackage{amsmath}
\usepackage{hyperref}
\usepackage{bigints}
\usepackage[utf8]{inputenc}
\usepackage{subcaption}
\usepackage{amsfonts,amssymb,amscd,amsthm,amsfonts,epsfig,enumerate,yfonts}
\usepackage{physics}
\usepackage[font=small,skip=0pt]{caption}
\usepackage{float}
\usepackage{multirow}
\usepackage{tikz}
\usepackage{ragged2e}
\usetikzlibrary{positioning}
\usepackage[font=small,skip=7pt]{caption}
\begin{document}
\title{Shadow Cast by the Kerr MOG Black Hole under the Influence of Plasma and Constraints from EHT Observations}
\author{Saira Yasmin$^{1}$}
\email{sairayasmeen555@gmail.com}
\author{Khadije Jafarzade$^{2}$}
\email{khadije.jafarzade@gmail.com}
\author{Mubasher Jamil$^{1}$}
\email{mjamil@sns.nust.edu.pk}
\affiliation{$^1$School of Natural Sciences, National University of Sciences and Technology (NUST), H-12, Islamabad 44000, Pakistan}
\affiliation{$^2$Department of Theoretical Physics, Faculty of Science, University of Mazandaran, P. O. Box 47416-95447, Babolsar, IRAN}
\date{\today}

\begin{abstract}
\begin{justify}
The study of black hole (BH) shadows provide crucial insights into the nature of strong gravitational effects and the intricate structure of the spacetime surrounding BHs.  In this paper, we explore the shadow of Kerr MOG  BH within a plasma environment, investigating how much the presence of plasma influences the characteristics of the observed shadow compared to those in vacuum conditions. Our analysis reveals that the shadow characteristics of M87* and Sgr A* are more compatible with event horizon telescope (EHT) observational data in nonhomogeneous plasma spacetime compared to homogeneous distributions. For small metric deformation parameter $\alpha$, the shadow aligns within $2\sigma$ uncertainty for homogeneous plasma and within $1\sigma$ for nonhomogeneous plasma. Next, we determine the energy emission rate for the Kerr MOG BH and analyze the influence of parameters $\alpha$, $k_o$,  $k_\theta$, and  $k_r$ on particle emissions in the BH vicinity. We further analyze the deflection angle in the presence of homogeneous and nonhomogeneous plasma profiles. The findings indicate notable differences from the vacuum scenario, underscoring the importance of accounting for plasma effects in studying light propagation around compact objects.
\end{justify}
\end{abstract}
\maketitle
\section{Introduction}
\begin{justify}
General theory of relativity $(\mathrm{GTR})$ is a well-established theory that successfully predicts many cosmological phenomena both in the solar system and the broader universe. Notably, it predicts the existence of BHs, which are among the most astonishing astronomical phenomena. However, $\mathrm{GTR}$ does not account for the interior of BHs or certain other astronomical observations.
One important example is the unusual behavior of galaxy rotation curves, discovered through the work of Zwicky, Rubin, and their colleagues \cite{zwicky1933redshift,rubin1980rotational}. Zwicky was the first to theorize and formally introduce the concept of an unknown form of matter, later termed dark matter. 
Rubin's observations revealed that Einstein's theory of relativity $(\mathrm{GTR})$ could not fully account for the rotation curves of spiral galaxies. This observation provides a key indication of the existence of dark matter.
\\
\indent However, dark matter has yet to be directly detected through experimental observation. It may be necessary to develop a theory that explains this phenomenon without relying on the concept of dark matter \cite{lee2017innermost}. Firstly, modified Newtonian dynamics (MOND) was proposed by Milgrom \cite{milgrom1983modification}, which modifies the "inverse-square" law of gravity in the scalar field framework of Newtonian gravity.
MOND fits certain observations, such as those within the weak-field approximation in systems like the solar system. However, it has limited ability to explain large-scale phenomena, such as the rotation curves of galaxies in the spiral region, from approximately $\sim0.2$ kpc to $\sim200$ kpc, and the velocity dispersion profiles of globular clusters \cite{bhattacharjee2014rotation,moffat2008testing}.
\\
\indent Moffat proposed a scalar–tensor–vector gravity (STVG) theory \cite{moffat2006scalar}, commonly referred to as modified gravity theory (MOG). MOG was developed to address the discrepancies between $\mathrm{GTR}$ at large scales and numerous astronomical observations, including galaxy dynamics, rotation curves, bullet clusters, the distribution of luminous matter in galaxies, and the accelerating expansion of the universe \cite{brownstein2007bullet}. Additionally, it provides a description of structure formation, the matter power spectrum, and the acoustic power spectrum of the cosmic microwave background (CMB) \cite{moffat2015black}.
This theory introduces new fields into $\mathrm{GTR}$, strengthening the gravitational field. Its action consists of the usual Einstein-Hilbert term associated with the metric tensor \(g_{\mu\nu}\), a massive vector field \(\phi_{\mu}\), and three scalar fields that represent the running values of the gravitational constant \(G\), the coupling constant \(\Omega\) (which determines the strength of the coupling between matter and the vector field), and the vector field’s mass \(\mu\) (which adjusts the range of the coupling) \cite{moffat2006scalar}. The scalar field \( G = G_N (1 + \alpha) \) represents the strength of the gravitational attraction, where \( G_N \) is Newton’s gravitational constant and \( \alpha \) is a dimensionless parameter of the theory.
\\
\indent Recently, the direct observation of gravitational waves by LIGO and Virgo from binary BH and neutron star systems \cite{Abbott2016observation,raidal2017observation} has sparked great excitement in the scientific community. These groundbreaking events have opened a new frontier for observational astronomy. As anticipated, the intrinsic properties of BHs will become increasingly clear with future detections of merging events and enhancements in signal to noise ratio. These observations may also offer us a pathway to differentiate BHs predicted by various theories of gravity \cite{cardenas2016testing,jafarzade2023rotating}.

There are other approaches to explore the nature of BHs, such as strong gravitational lensing and BH shadow analysis. In particular, these effects are significantly influenced by the mass of the BH. It is widely believed that a supermassive BH exists at the center of each galaxy \cite{akiyama2019first}. The immense mass of these BHs makes strong gravitational effects potentially observable. For example, the EHT, with its high angular resolution, can observe the shadow cast by the supermassive BH at the center of our galaxy using advanced imaging techniques. Therefore, studying the BH shadow is highly desirable \cite{hendi2023blackhole}.
BH shadows are formed by null geodesics within the region of strong gravitational influence. Generally, photons with high angular momentum originating from infinity are deflected back to infinity by the BH’s gravitational potential. In contrast, photons with lower angular momentum are pulled into the BH, creating a dark area visible to an observer at infinity. Between these two behaviors, photons with critical angular momentum will orbit the BH loop by loop, delineating the boundary of the shadow \cite{jafarzade2024study,jafarzade2021shadow}.
 For a non-rotating Schwarzschild BH, this shadow was first studied by Synge and Luminet \cite{Synge1966thescape,Luminet1979Image}.
 The shadow cast by a rotating Kerr BH was first investigated by Bardeen \cite{bardeen1972rotating} and was systematically introduced in \cite{chandrasekhar1998themathematical}. These studies reveal that a non-rotating BH produces a perfectly circular shadow, whereas the shadow of a rotating BH is elongated due to the frame-dragging effect. Recent studies have examined the shadows of rotating and non-rotating BHs in the presence of plasma \cite{pahlavon2024effect,hoshimov2024weak,atamurotov2023quantum,atamurotov2021axion} 
 \\
\indent Significantly, astronomical observables are essential in connecting theoretical studies with actual astronomical observations. Thus, investigating the properties of BHs within MOG theory is of great interest. In \cite{moffat2015black,guo2018observational}, the authors presented a preliminary image of the shadow cast by the MOG BH. In this paper, we aim to explore the observable characteristics of the Kerr MOG BH shadow under the influence of plasma. These findings will offer detailed insights into the properties of MOG theory and provide a potential means to differentiate it from $\mathrm{GTR}$.
 \\
\indent  The effects of plasma on light propagation have been investigated since the 1960s. In 1966, John Ston and Muhleman explored the impact of the electronic plasma in the solar corona on the time delay of radio frequencies under the gravitational field of the sun, using the plasma and gravitational refractive indices to derive a weak-field-approximation \cite{muhleman1966radio}.
   Using light propagation studies on different space missions such as Viking, Mariner 6 and 7, and the Cassini mission, various analyses centered on the solar wind and the electron density profile in its outer corona were also conducted. In 1980, Ehlers and Breuer enacted a challenging derivation of a Hamiltonian for light rays consisting of magnetized plasma in a curved spacetime \cite{breuer1980propagation}. In literature, we can find recent work analyzing the influence of plasma on the propagation of light rays in different astrophysical situations \cite{pahlavon2024effect}.
   \\
\indent The EHT group recently has successfully caught a polarized light around the M87* and Sgr A* \cite{EventHorizonTelescope:2021srq}. The orientation of light waves is apparent in these polarized images, and it is influenced by the magnetic field generated by the plasma revolving around the BHs. The results from these polarized images not only assure the presence of plasma around BHs but also give observations into the mechanisms that lead to their evolution and behavior \cite{akiyama2024first}. 
This paper investigates Kerr MOG BH and its implications for the EHT observations of M87* and Sgr A*, focusing on how the deformation parameter $\alpha$ modifies the BH’s shadow and the constraints derivable from EHT data in the presence of non-magnetized, pressureless plasma. By comparing the projections of the Kerr MOG model with observed shadows, we aim to constrain $\alpha$, providing a potential test for deviations from $\mathrm{GTR}$. The shadow of M87* is consistent with the predictions of $\mathrm{GTR}$ for a Kerr BH; however, uncertainties in spin measurements and deviations in quadrupole moments allow for potential modifications \cite{akiyama2019first,cardoso2019testing} due to $\alpha$ and plasma effects.  By incorporating plasma effects and the deformation parameter $\alpha$, this work evaluates the feasibility of distinguishing Kerr MOG BHs and constraining their parameters through EHT observations.
\\
\indent Another key aspect of this study is deriving the deflection angle of light near the Kerr MOG BH, considering both homogeneous and nonhomogeneous plasma spacetime. In addition, for most astronomical situations, the influence of plasma on light rays can be neglected; however, this is not the case for light rays in the radio frequency range. A well-known example is the effects of the solar corona, modeled as a non-magnetized and pressureless plasma, on the time delay \cite{muhleman1966radio} and deflection angle when light rays propagate near the Sun \cite{muhleman1970radio}. Later, Perlick conducted a detailed investigation into the influence of a nonhomogeneous plasma on light deflection in the Schwarzschild spacetime and within the equatorial plane of the Kerr spacetime \cite{perlick2000ray}. Here, we investigate the deflection angle of light in the Kerr MOG spacetime, considering a non-magnetized and pressureless plasma. The influence of plasma on light propagation is particularly significant in the context of MOG, making this analysis crucial for understanding such effects. The analysis focuses on the impacts of the plasma properties and the deformation parameter $\alpha$ on the deflection angle.
   \\
\indent The organization of the paper is as follows. In Sec.~II, we summarize the metric described in \cite{moffat2015black} and in Sec.~II A, we review the Hamiltonian formalism for light rays in a non-magnetized, pressureless plasma within a general-relativistic spacetime and specialize the relevant equations to the Kerr MOG metric.
In Sec.~II B, we derive the necessary and sufficient condition on the plasma electron density that ensures the separability of the Hamilton-Jacobi $(\mathrm{HJ})$ equation for light rays \cite{perlick2017light}, thereby guaranteeing the existence of a Carter constant. In Sec.~II C, we determine the photon region for a plasma density around a Kerr MOG BH that satisfies this separability condition. In Sec.~III, we develop the general formulas needed to parameterize the contour curve of the BH shadow in pressureless, non-magnetic plasma environments adapted to the Kerr MOG metric. In Sec.~III A, we begin with considerations of BH shadows in homogeneous plasma environments, and then we examine how the shadows and photon regions are influenced by various parameters of the BH and plasma profiles \cite{perlick2017light} in  Sec.~III B and  Sec.~III C. In Sec.~IV, we analyze the energy emission rate and investigate the effects of the MOG parameter and plasma density on particle emission. In Sec.~V, after examining Hawking radiation in Sec.~IV, we explore the deviation of light in Kerr MOG spacetime surrounded by non-magnetized and pressureless plasma. Sec.~VI focuses on constraining the plasma parameters by incorporating the observational data obtained from M87* and Sgr A*. We conclude with final remarks in Sec.~VII.
Our conventions are as follows: we employ the summation convention for Greek indices, which take the values 0, 1, 2, and 3. Our choice of signature is \((-,+,+,+)\). We raise and lower Greek indices using the spacetime metric and we will employ the convention of setting \( G_N = c = 1 \), where \(G_N\) represents the gravitational constant, and \(c\) denotes the vacuum speed of light.
\end{justify}
\section{Null Geodesics in Kerr MOG spacetime}
\begin{justify}
The Kerr MOG BH is  axisymmetric solution of
the gravitational field equations. The spacetime geometry of a rotating Kerr MOG BH is governed by its angular momentum $a$, mass $M$, and a deformation parameter $\alpha$ is given by (in Boyer Lindquist coordinates) \cite{moffat2015black}
 \begin{align}\label{kmogmetric}
d s^{2}= & -\frac{\Delta-a^{2} \sin ^{2} \theta}{\rho^{2}} d t^{2}+\sin ^{2} \theta\left(\frac{\left(r^{2}+a^{2}\right)^{2}-\Delta a^{2} \sin ^{2} \theta}{\rho^{2}}\right) d \phi^{2}  \nonumber \\
& -2 a \sin ^{2} \theta\left(\frac{r^{2}+a^{2}-\Delta}{\rho^{2}}\right) d t d \phi+\frac{\rho^{2}}{\Delta} d r^{2}+\rho^{2} d \theta^{2} ,
\end{align}
\begin{align}
\Delta=r^{2}-2 G M r+a^{2}+\alpha G_{\mathrm{N}} G M^{2}, \quad \rho^{2}=r^{2}+a^{2} \cos ^{2} \theta, 
\end{align}
where, $G =G_N(1+\alpha)$ is an enhanced gravitational constant with the contribution of Newton's gravitational constant $G_N$ and the deformation rate $\alpha$. The Newtonian mass $M$ and the ADM mass $\mathcal{M}$ are related with
$\mathcal{M}=(1+\alpha) M$. The Kerr MOG metric has two horizons known as the outer horizon \( (r_+) \) and the inner horizon \( (r_-) \), similar to a Kerr BH determined by $\Delta=0$, namely
\begin{align}
 r^2-2 \mathcal{M} r +a^2+\frac{\alpha}{1+\alpha }\mathcal{M}^2=0.
\end{align}
The two horizons are also referred to as the event horizon and the Cauchy horizon, respectively. They are given by \cite{lee2017innermost}
\begin{equation}
r_{\pm} = G_{\mathrm{N}} \mathcal{M} \pm \sqrt{\frac{G_{\mathrm{N}}^2 \mathcal{M}^2}{1+\alpha} - a^2}.
\end{equation}
\end{justify}
\subsection{Hamiltonian formulation for light rays in plasma on Kerr MOG spacetime}
\begin{justify}
A Hamiltonian for light propagation in a non-magnetized pressureless plasma is defined as \cite{perlick2000ray}
 \begin{equation}
\mathcal{H}(x, p)=\frac{1}{2}\left(g^{\mu \nu}(x) p_{{\mu}} p_{\nu}+\omega_\text{p}(x)^{2}\right).
\end{equation}
where $x =\left(x^{0}, x^{1}, x^{2}, x^{3}\right)$ denote spacetime coordinates and $ p =\left(p_{0}, p_{1}, p_{2}, p_{3}\right)$ are canonical momentum coordinates. The plasma frequency $\omega_\text{p}$ of the medium determined as
\begin{equation}
\omega_\text{p}(x)=\frac{4 \pi e^{2}}{m} N_\text{{e}}(x),
\end{equation}
in which $m$ and $N_\text{e}$ are the mass of the electron and electron number density respectively, and $e$ denotes the electric charge. In the case of a vacuum, the plasma frequency is defined as \( \omega_\text{p}(x) = 0\). A plasma is a dispersive medium with an index of refraction $n$ given by \cite{perlick2017light}
\begin{align}\label{h1}
 n(x,\omega)^2&=1-\frac{\omega_\text{p}(x)^2}{\omega(x)^2},
\end{align}
which depends on $\omega(x)$ (photon frequency) concerning plasma frequency.
In this medium light propagation is possible only if \cite{liu2016electromagnetic}
\begin{align}\label{wx<wp}
    \omega_\text{p}(x)\le\omega(x),
\end{align}
which assures a nonnegative and real index of refraction. With the metric coefficients defined by the Kerr MOG metric (\ref{kmogmetric}), the Hamiltonian in (\ref{h1}) takes the form
 \begin{align}\label{hamil final1}
  \mathcal{H} & = \frac{1}{2\rho^2}\left[\frac{-1}{\Delta }\left((r^2+a^2){p_{\text{t}}}+a p_\phi \right)^2 \right.+\left(\frac{p_\phi}{\sin{\theta}}+a \sin{\theta}{p_{\text{t}}}\right)^2+\Delta p_{\text{r}}^2 +  p_\theta^2
  + \rho^2 \omega_\text{p}(r, \theta)^2 \bigg].
\end{align}
If we suppose that $\omega_\text{p}$ only depends on $r$ and $\theta$, than $\frac{\partial H}{\partial t}=0$ and $\frac{\partial H}{\partial \phi}=0$ which implies that $p_{\text{t}}= -E = -\omega_0$ and $p_\phi=L_\text{z}$ are constant of motions. $p_\phi$ is angular momentum and $\omega_0$ is the frequency of photons as measured by a stationary observer at infinity \cite{perlick2017light}.
\end{justify}
\subsection{Separation of Hamilton-Jacobi equation for light rays in Kerr MOG spacetime with plasma}
\begin{justify}
According to (\ref{hamil final1}), the Hamilton-Jacobi ($\mathrm{HJ}$) equation is \cite{perlick2017light}
\begin{align}\label{h(x,pSpx)}
0 = \mathcal{H} \left(x,\frac{\partial S}{\partial x}\right).  
\end{align}
Using the expressions above, (\ref{h(x,pSpx)}) transforms into the following
\begin{align}\label{hj0}
0= & -\frac{1}{\Delta}\left(a \frac{\partial S}{\partial \phi}+\left(r^{2}+a^{2}\right)\frac{\partial S}{\partial t}\right)^{2} +\left(\frac{1}{\sin \theta} \frac{\partial S}{\partial \phi}+ a\sin \theta \frac{\partial S}{\partial t}\right)^{2}\nonumber \\
& +\left(\frac{\partial S}{\partial \theta}\right)^{2}+\Delta\left(\frac{\partial S}{\partial r}\right)^{2}+\rho^{2}{\omega_p^{2}}.
\end{align}
Here, we have constants of motion \( p_t = \frac{\partial S}{\partial t} \) and \( p_\phi = \frac{\partial S}{\partial \phi} \), corresponding to the conserved energy and angular momentum of the light ray, respectively. Now, by using separation ansatz \cite{houchmandzadeh2020hamilton}
\begin{align}\label{sep ansatz plasma1}
S(t, r,\theta,\phi)&=- E t+S_{r}(r)+S_{\theta}(\theta)+ L_\text{z} \phi.
\end{align}
By substituting the $S_{\theta}^{\prime}(\theta)=\frac{\partial S}{\partial \theta}$ and $ S_{r}^{\prime}(r)=\frac{\partial S}{\partial r}$  into (\ref{hj0}), we obtain the following
\begin{align}\label{seprathamil}
0=- & \frac{1}{\Delta}\left(a L_\text{z}-\left(r^{2}+a^{2}\right) E\right)^{2}+\left(\frac{L_\text{z}}{\sin {\theta}}-a E\sin{\theta}\right)^{2} +\Delta S_{r}^{\prime}(r)^{2}+S_{\theta}^{\prime}(\theta)^{2} \nonumber \\ &+{\omega_p^{2}}\left(r^{2}+a^{2} \cos ^{2} \theta\right).
\end{align}
The $\mathrm{HJ}$ equation, \(\mathcal{ H} = 0 \), will be separable in the variables \( r \) and \( \theta \) only if the plasma distribution takes the form \cite{perlick2017light}
\begin{equation} \label{plasma density}
\omega_\text{p}(r, \theta)^{2}=\frac{f_\text{r}(r)+f_{\theta}(\theta)}{r^{2}+a^{2} \cos ^{2} \theta},
\end{equation}
with some functions $f_\text{r}(r)$ and $f_{\theta}(\theta)$.
Thus, using equation (\ref{plasma density}), we can ensure that the equations of motion are fully integrable. The $\mathrm{HJ}$ equation then takes the form
\begin{align}
-\Delta S_{\text{r}}^{\prime}(r)-\frac{2a(r^2+a^2-\Delta)}{\Delta}E L_\text{z}+\frac{(r^2+a^2)^2 E^2}{\Delta}+\frac{a^2}{\Delta}L_\text{z}^2 -a^2 E^2 - L_\text{z}^2- f_{\text{r}}(r) 
\\\nonumber =  S_{\theta}^{\prime}(\theta)^{2} +  \frac{L_\text{z}^2 \cos^2 \theta}{\sin^2 \theta} - a^2E^2 \cos^2 \theta + f_{\theta}(\theta) =: \mathcal{C},
\end{align}
where $\mathcal{C}$ is the Carter constant. With $S_{\theta}^{\prime}(\theta)= p_{\theta}$ and   $ S_{r}^{\prime}(r)= p_\text{r}$, we can express the components of the 4-momentum as
\begin{align}\label{pthetapr}
p_{\theta}^{2} &= \mathcal{C} - \left( \frac{L_\text{z}^2}{\sin^2 \theta} - a^2E^2 \right)\cos^2 \theta - f_{\theta}(\theta),   \\
\phantom{p_{\theta}^{2}} p_{\text{r}}^{2} &= \frac{1}{\Delta }\left(-\mathcal{C} +\frac{(r^2+a^2)^2 E^2}{\Delta} - \frac{2a(r^2+a^2-\Delta)}{\Delta}E L_\text{z}+\frac{a^2}{\Delta}L_\text{z}^2-a^2 E^2-L_\text{z}^2-f_{\text{r}}(r)\right).
\end{align}
The insertion of equation (\ref{pthetapr}) into Hamilton's equations $\dot{x}^\mu=\frac{\partial H}{\partial p_\mu}$ for $x^\mu= (t,r,\theta,\phi)$ , produces the following first-order equations of motion for the photon given as
\begin{align}
\rho^4 \dot{\theta}^2 &= \mathcal{C} - \left( \frac{L_\text{z}^2}{\sin^2 \theta} - a^2E^2 \right)\cos^2 \theta - f_{\theta}(\theta), \label{thetadot2} \\[1ex]
\rho^4 \dot{r}^2 &=\left( \left( r^{2} + a^{2} \right) E - a L_{\text{z}} \right)^{2}-\Delta\left(\mathcal{C}+(L_\text{z}-a E)^2\right) - f_{\text{r}}(r) \Delta  =: \mathcal{R}(r), \label{rdot2} \\[1ex]
\rho^2 \dot{t} & = \frac{1}{\Delta} \left[\left((r^2 + a^2)^2-\Delta a^2 \sin^2\theta)E \right. \right.- \left.  \left(2  \mathcal{M} r -  a \frac{\alpha} {(1+\alpha)} \mathcal{M}^2\right) L_\text{z} \right],\label{final_eq} \\[1ex]
\rho^2 \dot{\phi} &= \frac{1}{\Delta} \Bigg[ 
   a \left( 2  \mathcal{M} r  -   \frac{\alpha} {(1+\alpha)} \mathcal{M}^2  \right) E 
   + \left( \rho^2 - 2  \mathcal{M} r + \frac{\alpha} {(1+\alpha)} \mathcal{M}^2 \right) L_\text{z} \csc^2 \theta \Bigg]. \label{phidot2}
\end{align}
 The radial equation of motion (\ref{rdot2}) can be rewritten in terms of effective potential $\mathcal{R}(r)$ as
\begin{align}
      \mathcal{R}(r)&= -\zeta \Delta+(r^2+a^2)^2-2a(r^2+a^2-\Delta)\varphi+a^2 \varphi^2-a^2 \Delta-\Delta \varphi^2-\frac{f_r(r)}{E^2}\Delta.
  \end{align}
Here we introduced the impact parameters
  $\varphi=\frac{\mathrm{L}_z}{\mathrm{E}}$ and  $\zeta=\frac{\mathcal{C}}{\mathrm{E}^2}$, a third conserved quantity $\mathcal{C}$ derived by the separability of $\mathrm{HJ}$ equation.
To conclude, condition (\ref{plasma density}) is both necessary and sufficient for the existence of the Carter constant, which ensures the full integrability of the equations of motion.
\end{justify}
\subsection{Unstable circular photon rbits}
\begin{justify}
The silhouette of a BH is determined by the unstable circular photon orbits with a fixed radial coordinate \( r= r_\text{p} \), which must satisfy the conditions, given as $
\mathcal{R}(r) = 0, ~ \frac{d\mathcal{R}}{dr} = 0$ \cite{teo2021spherical}. 
From the above two conditions, the following can be derived 
\begin{align}
    \varphi &= \frac{-2 a r \Delta + a^3 \Delta' + a r^2 \Delta' - \sqrt{4 a^2 r^2 \Delta^2 - a^2 \Delta^2 \tilde{f}_{\text{r}}'(r) \Delta'}}{a^2 \Delta'}, 
    \label{varphi} \\
   \zeta &= \frac{1}{a^2 \Delta'} \Bigg( 
        4 r^3 \Delta 
        - a^2 \Delta \tilde{f}_\text{r}'(r) 
        + \Delta^2 \tilde{f}_\text{r}'(r) 
        + \frac{8 a^2 r^2 \Delta}{\Delta'} 
        - \frac{8 r^2 \Delta^2}{\Delta'} 
        - a^2 \tilde{f}_\text{r}(r) \Delta' 
        - r^4 \Delta'
        \nonumber \\  
        &\quad + \sqrt{a^2 \Delta^2 \big(4 r^2 - \tilde{f}_\text{r}'(r) \Delta'\big)} \Bigg( 
            \frac{2 r^2}{a} 
            + \frac{4 a r}{\Delta'} 
            - \frac{4 r \Delta}{a \Delta'}
        \Bigg) 
    \Bigg)
 \label{zeta},
\end{align}
where \( \tilde{f}_\text{r} \equiv \frac{f_r(r)}{\omega_0^2} \), and we have introduced the photon energy \( E = \omega_0 \). Certain critical values of the parameters \( \varphi \) and \( \zeta \) can be characterized, and under small perturbations, these critical values lead to either escape or plunge orbits for photons. In this regard, it is important to note that the outline of the shadow is primarily dependent on the critical impact parameters. Since \( \zeta = 0 \) at the equatorial plane, the roots of (\ref{varphi}) and (\ref{zeta}) determine the critical orbits of the photons \( r_{\text{p}} \) in Kerr MOG spacetime, respectively, at that plane.
We can obtain the apparent shape of the shadow seen by the observer if we consider the celestial coordinates $X$ and $Y$, which are the coordinates of the observer's sky. The general expression to find celestial coordinates $X$ and $Y$ is given by 
\cite{bardeen1972rotating}
\end{justify}
\begin{align}\label{alpha}
    X &=\lim_{r_0 \to \infty}\left(-r_0^2 \sin \theta_0\frac{d\Psi}{dr} |_{(r_{0},\mathrm{\theta_0)}} \right),\nonumber \\
    Y &=\lim_{r_0 \to \infty}\left(r_0^2 \frac{d\Phi}{dr} |_{(r_{0},\mathrm{\theta_0)}} \right).
\end{align}
With the presence of the plasma, the celestial coordinates are modified as follows \cite{perlick2017light}
\begin{align}\label{XY}
    X &=-\varphi \csc \theta_0, \nonumber \\
    Y&= \sqrt{\zeta +a^2\cos^2\theta_0-\varphi^2 \cot^2 \theta_0-\frac{f_\theta(\theta)}{\omega_0^2}}.
\end{align}
\section{The Shadow in Plasma-Filled Environment}
\begin{figure}[H] 
     \centering
      {{\includegraphics[height=7cm,width=7cm]{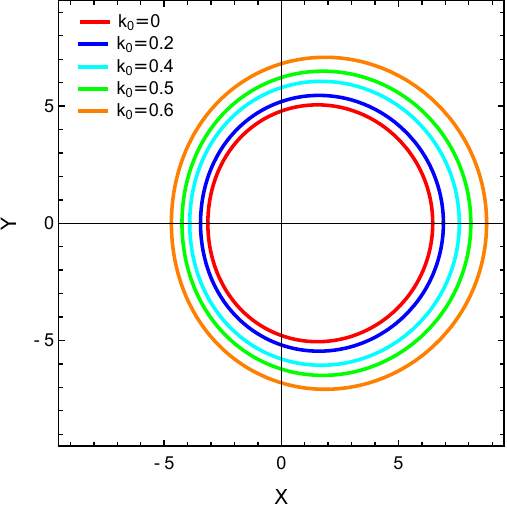}}}
       {{\includegraphics[height=7cm,width=7cm]{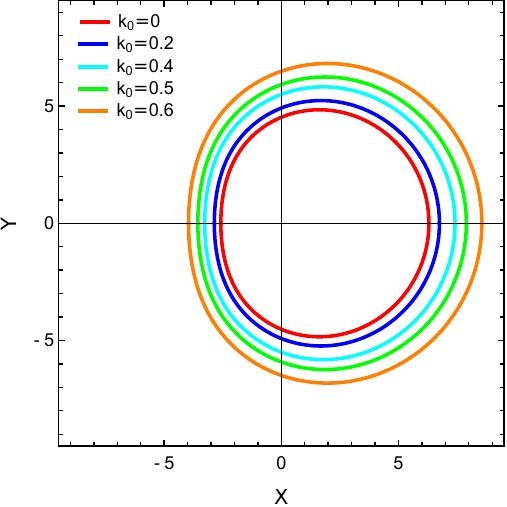}}}
    
       \caption{\label{figab} Shadows of a black hole for varying homogeneous plasma parameter $k_0$ and deformation parameters $\alpha$. The observer's angle is fixed at $\theta_0 = \pi/2$, with spin $a = 0.75$. The left plot illustrates the shadow for $\alpha = 0.19$, while the right plot corresponds to $\alpha = 0.6$.
    }
    \end{figure}

\begin{figure}[H] 
     \centering
      {{\includegraphics[height=7cm,width=7cm]{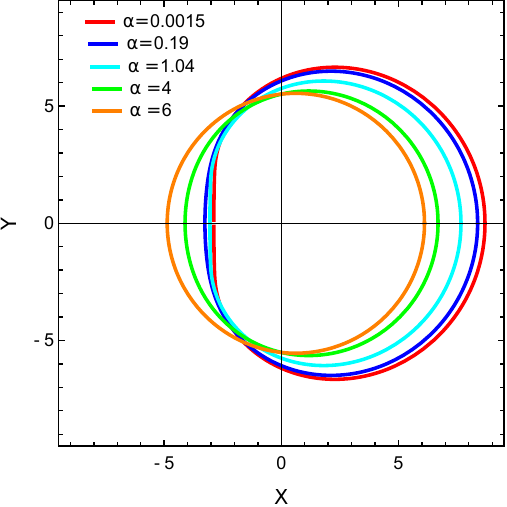}}}
       {{\includegraphics[height=7cm,width=7cm]{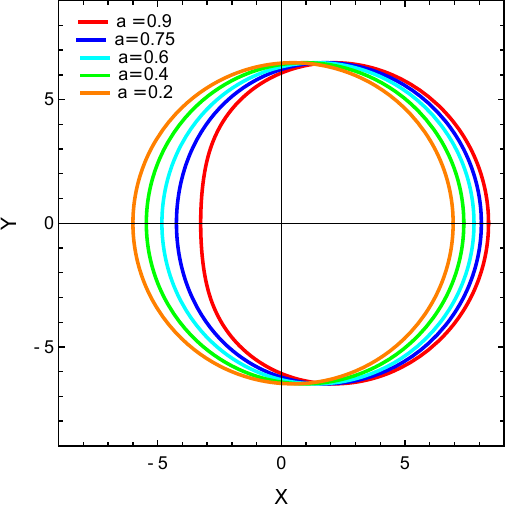}}}
       \caption{\label{figa}Shadows of a black hole with varying spins and deformation parameter $\alpha$. In the right plot, $\alpha = 0.19$, while the left plot presents the shadow with different spins
$a = 0.2$ (leftmost, orange), $a = 0.4$ (green), $a = 0.7$ (cyan), $a = 0.9$ (blue), and $a=0.999$ (red) with deformation parameter $\alpha$. The observer's angle is fixed at $\theta_0 = \pi/2$, and the plasma parameter is set to $k_0 = 0.5$.}    
    \end{figure}

\begin{figure}[H] 
     \centering
      {{\includegraphics[height=6.9cm,width=6.9cm]{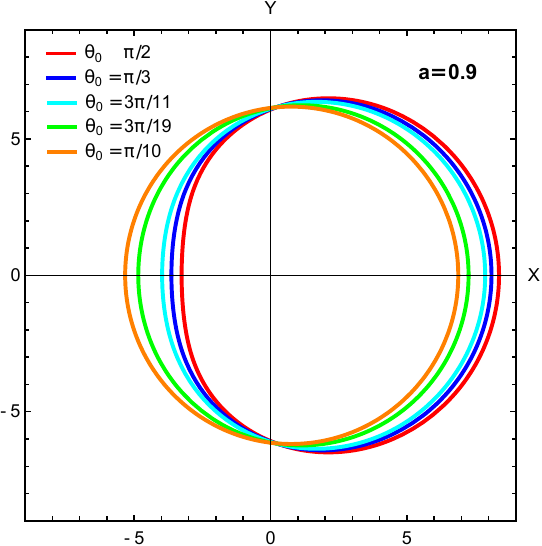}}}
      {{\includegraphics[height=6.9cm,width=6.9cm]{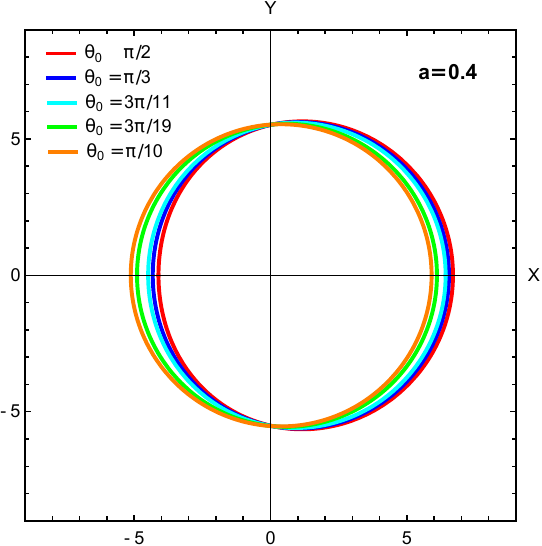}}}
     \\
       \caption{\label{figb}Shadows for different angular positions \( \theta_0 \) of the observer.  The deformation parameter is set to \( \alpha = 0.0015\) in the left plot and \( \alpha = 4\) in the right plot and \( k_0 = 0.5 \). 
}
       \end{figure}

\begin{figure} [H] 
     \centering
     \includegraphics[height=9cm,width=9cm]{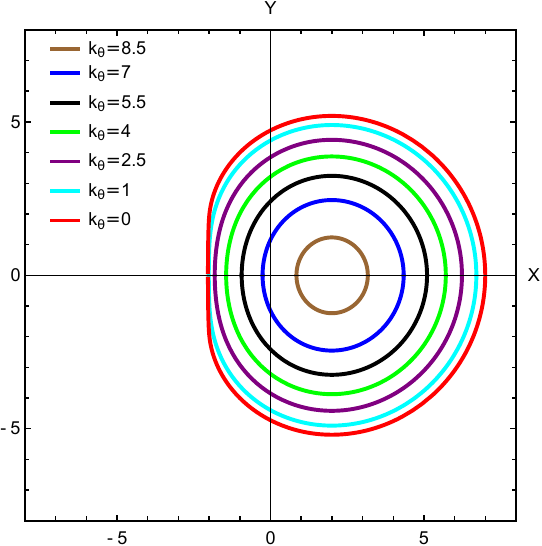}
     \caption{\label{figc}This figure shows the shadow for an observer at $\theta_0=\frac{\pi}{2}$ with the  spin $a=0.999$  and deformation parameter $\alpha=0.0015$. When the forbidden region reaches the observer position the shadow has shrunk to a point and it vanished at $k_\theta \approx 9.1$}
     \end{figure}

\begin{figure} [H] 
     \centering
     \includegraphics[height=9cm,width=9cm]{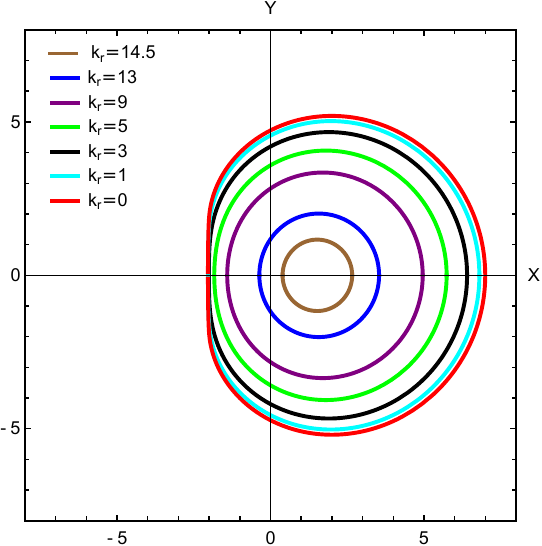}
     \caption{\label{figd}This figure shows the shadow for an observer at infinity with angular position $\theta_0=\frac{\pi}{2}$ with the spin $a=0.999$  and deformation parameter $\alpha=0.0015$. At  $ k_r \approx 15.25$ it has vanished.}
     \end{figure}

\begin{justify}
In this section, we will examine various plasma distributions and their impact on the formation of the BH shadow.  We will analyze different plasma distributions and how they affect the formation of the black hole shadow. To ensure the separability of the HJ equations, the proposed distributions must satisfy (\ref{plasma density}). Our objective here is to define the functions \( f_\text{r}(r) \) and \( f_\theta(\theta) \). This will be achieved by mainly considering the plasma distributions discussed in \cite{perlick2017light}.
\end{justify}
\begin{table}
 \centering
 \caption{\label{taba}Corresponding to different values of the spin parameter $a$, the deformation parameter $\alpha$ range is shown for the Kerr MOG BH. Here we have set mass parameter as $\mathcal{M}=1$ \cite{sheoran2018mass}.}
\begin{tabular}{lll} 
\hline
No.  & \quad \quad $a/\mathcal{M}$ & \quad  \quad Range of $\alpha$ \\
\hline
1 &\quad  \quad 0.3 & \quad  \quad $0 \leq \alpha \leq 10.111$ \\
2 & \quad  \quad 0.4 &  \quad  \quad$0 \leq \alpha \leq 5.250$ \\
3 &  \quad  \quad 0.5 &  \quad  \quad$0 \leq \alpha \leq 3.0$ \\
4 &  \quad  \quad 0.6 &  \quad  \quad $0 \leq \alpha \leq 1.777$ \\
5 &  \quad  \quad 0.7 &  \quad  \quad$0 \leq \alpha \leq 1.040$ \\
6 &  \quad  \quad 0.8 &  \quad  \quad $0 \leq \alpha \leq 0.562$ \\
7 &  \quad  \quad 0.9 &  \quad  \quad $0 \leq \alpha \leq 0.234$ \\
8 &  \quad  \quad 0.99 & \quad  \quad  $0 \leq \alpha \leq 0.02$ \\
9 &  \quad  \quad 0.999 & \quad  \quad $0 \leq \alpha \leq 0.002$ \\
\hline
\end{tabular}
\end{table}
 \begin{justify}

\subsection{Homogeneous plasma (${\omega_p^{2}}(r, \theta)=$constant)} 
\begin{justify}
 In this case, we observe the shadow of the Kerr MOG BH in the homogeneous plasma environment. For this, we consider the plasma distribution given as
 \begin{align}
f_\text{r}(r)=k_0 \omega_\text{{0}}^{2} r^{2} \quad f_{\theta}(\theta)=k_0 \omega_\text{{0}}^{2}a^{2} \cos ^{2} \theta, 
 \end{align}
 where \( k_0 = \frac{\omega_\text{p}^2}{\omega_0^2} \)
 represents the homogeneous plasma parameter which describes the homogeneous plasma distribution. This parameter must lie within the range  (0,1) to satisfy (\ref{wx<wp}) that describes the plasma frequency does not exceed the photon frequency. This condition is qualitatively the same as without plasma.
However, it is important to note that the shadow boundaries are not defined by (\ref{XY}) for a homogeneous plasma profile. Instead, the celestial coordinates in this case are determined by solving (\ref{alpha}) along with the geodesic equations given by \cite{kumar2024observational}
\begin{align}
    X &=-\frac{\varphi}{ \sin \theta_0\sqrt{1-k_0}}, \nonumber \\
    Y&= \frac{\sqrt{\zeta +a^2\cos^2\theta_0-\varphi^2 \cot^2 \theta_0- k_0 a^2 
 \cos^2\theta}}{\sqrt{1-k_0}},
\end{align}
In this case Fig.~(\ref{figab}), the shadow exhibits distinct shapes for different values of $k_0$, while the effect of $\alpha$ becomes apparent as its value increases within the range specified in the table, causing the shadow to appear more deformed. Fig.~(\ref{figa}) depicts the shadows of a Kerr MOG BH for different deformation parameters within the specified range and varying spin parameters, respectively. In Fig.~(\ref{figa}) (right panel), we can see that the spin of a black hole influences its shadow in three ways. First, as the spin increases, the shadow becomes smaller in size, reducing its horizontal and vertical diameters. Second, the shadow's shape transitions from nearly circular to a distinct D-shape, with the flattened side aligned vertically due to co-rotating photon orbits. Third, the shadow shifts to the right, with this displacement becoming more pronounced at higher spin. In Fig.~(\ref{figa})  (left panel), we can see that as the value of $\alpha$ increases, with lower values of spin, the shadow becomes circular and its size decreases.
 Fig.~(\ref{figb}) illustrates how the angle \( \theta_0 \) between the position of the observer and the rotation axis of BH affects the shape, size, and position of the shadow. Observers near the rotation axis (\( \theta_0 \rightarrow 0 \)) will see a smaller, rounder, and more centered shadow, while observers near the equatorial plane (\( \theta_0 \rightarrow \pi/2 \)) will observe a displaced shadow with a larger vertical diameter and a D-shaped appearance. Here, we can see how the Kerr MOG spacetime affects the shape of the BH shadow and the role that the parameter \( \alpha \) plays in this relationship. As \( \alpha \) approaches zero, the metric converges to the standard Kerr spacetime. However, with an increase in $a$ with deformation parameter $\alpha$ and $k_o$, the shadows exhibit distinct evolution patterns that are governed by the properties of the Kerr MOG metric in the presence of homogeneous plasma.
 \end{justify}
 \subsection{Plasma distribution with $f_r(r)=0$}
 \begin{justify}
Now, we investigate the behavior of Kerr MOG BH shadow in an inhomogeneous plasma environment. Here we choose 
 $f_\text{r}(r)=0$ and $f_{\theta}(\theta)=k_\theta \omega_\text{{0}}^2 {M}^2 (1+2 \sin ^2\theta)$ and plasma density resulting as
 \begin{align}
\frac{{\omega_p^{2}}}{\omega_0^2}=\frac{k_\theta \mathcal{M}^2 (1+2 \sin ^2\theta)}{(1+\alpha)^2(r^{2}+a^{2} \cos ^{2} \theta)},
\end{align}
 where $k_\theta$ is the latitudinal plasma parameter.
  \end{justify}
    \end{justify}
The parameter $k_\theta$ governs the latitude-dependent component of the plasma distribution, which remains significant near the black hole.  Although the overall density follows a behavior $\sim 1/r^2$ at large $r$, latitudinal variations influence the photon motion and the shadow structure.
 The celestial coordinates for this case are given by (\ref{XY}). The shadows corresponding to this plasma density are presented in Fig.~(\ref{figc}). A forbidden region emerges around the equatorial plane, dividing the photon region into two disconnected areas. As the plasma frequency increases, this forbidden region expands until it completely encircles the BH. Low-energy photons cannot penetrate this region and are instead deflected.       
    \begin{justify}
   A forbidden region emerges around the equatorial plane, dividing the photon region into two disconnected areas. As the plasma frequency increases, this forbidden region expands until it completely encircles the BH. Low-energy photons cannot penetrate this region and are instead deflected.  Consequently, observers located close to the equatorial plane (\( \theta_0 = \pi/2 \)) are the first to lose sight of the shadow. The resulting view will be entirely bright for observers beyond the forbidden region and completely dark for those situated between the forbidden region and the BH.
       In Fig. (\ref{figc}), the shadow becomes invisible for certain plasma frequencies, leaving a completely illuminated sky. By examining the photon region, it is observed that a specific critical frequency ratio $\frac{\omega_\text{p}^2}{\omega_0^2}$ exists. Beyond this threshold, a forbidden zone appears where the propagation condition (\ref{wx<wp}) is no longer met. As the plasma frequency increases, the forbidden region expands until it fully encloses the BH.
        \subsection{Plasma distribution with $f_\theta(\theta)=0$} 
        \end{justify}
     \begin{justify}
     An interesting example is the case where the plasma frequency depends on \( r \) and \( \theta \), corresponding to the mass density of dust at rest at infinity. Shapiro \cite{Shapiro1974accretion} provided the analytical form of this matter distribution, even considering a non-zero pressure. When specialized to the pressureless 
         case, the mass density is found to be independent of \( \theta \) and proportional to \( r^{-3/2} \).  In this scenario, the separability condition (\ref{plasma density}) is not satisfied in a Kerr MOG spacetime with \( a \neq 0 \), which means that our calculation approach cannot be directly applied.
\end{justify}
     However, analytical formulas can still be used if we assume an additional \( \theta \) dependence on the plasma frequency. {The plasma density in our model retains a $\theta$ dependence due to the term $\rho^2 = r^2 + a^2 \cos^2 \theta$ in the denominator, which naturally arises from the metric structure. This ensures compatibility with the separability of the $\mathrm{HJ}$ equation, allowing us to proceed with our calculations.} To satisfy (\ref{plasma density}), requires adjustments to the profile, resulting in 
     \begin{align}
   \frac{{\omega_p^{2}}}{\omega_0^2}=\frac{k_r\sqrt{\mathcal{M}^3 r}}{(1+\alpha)^\frac{3}{2}(r^{2}+a^{2} \cos ^{2} \theta)}. 
\end{align}
where $k_r$ is radial plasma distribution parameter. The celestial coordinates for this case can be represented by Equation (\ref{XY}), given as follows \cite{bardeen1972rotating}
  \begin{justify}
     \begin{align}
    X &=-\varphi \csc \theta_0, \nonumber \\
    Y&= \sqrt{\left(\zeta +a^2\cos^2\theta_0-\varphi^2 \cot^2 \theta_0\right)}.
\end{align}
This profile approaches Rogers’ profile as the distance from the BH increases.
      The shadows obtained for this plasma profile are shown in Fig.~(\ref{figd}).  
     Here, we can observe that by increasing the ratio $\frac{\omega_\text{p}^2}{\omega_0^2}$ expands the forbidden region until it entirely encloses the BH at a critical value of $\frac{\omega_\text{p}^2}{\omega_0^2}$.
 At this point, the shadow vanishes for all observers, leading to a completely bright sky, as in Fig.~(\ref{figd}) with $k_r \approx 15.25$ shadow vanishes. 
The ratio $\frac{\omega_p^{2}}{\omega_0^{2}}$ in this case, also varies with the MOG parameter $\alpha$. As the values of the spin parameter are increased within the deformation parameter $\alpha$, the ratio $\frac{{\omega_p^{2}}}{\omega_0^{2}}$ also increases. With the increase in plasma frequency, the effects of the plasma become more significant, counteracting the gravitational effects \cite{briozzo2023analytical}. This occurs because, for over-dense plasma distributions, the plasma behaves as a repulsive medium, unlike the attractive nature of the gravitational field. Consequently, the shadow shrinks as the plasma frequencies rise.
\end{justify}
Next, we compare the behavior of the Kerr MOG black hole by considering both homogeneous and inhomogeneous plasma distributions.
 \begin{figure} [H] 
     \centering
     \includegraphics[height=6.9cm,width=6.9cm]{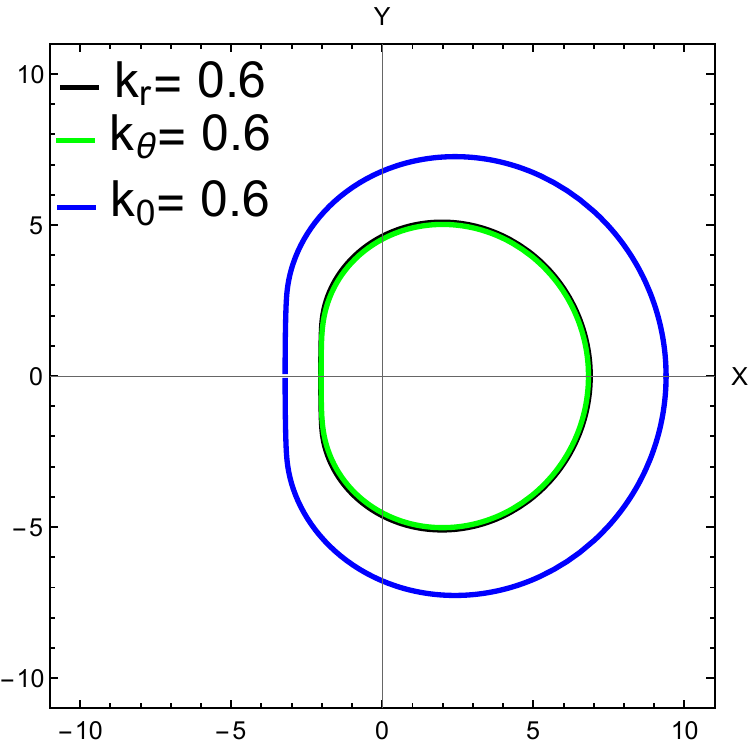}
     {{\includegraphics[height=6.9cm,width=6.9cm]{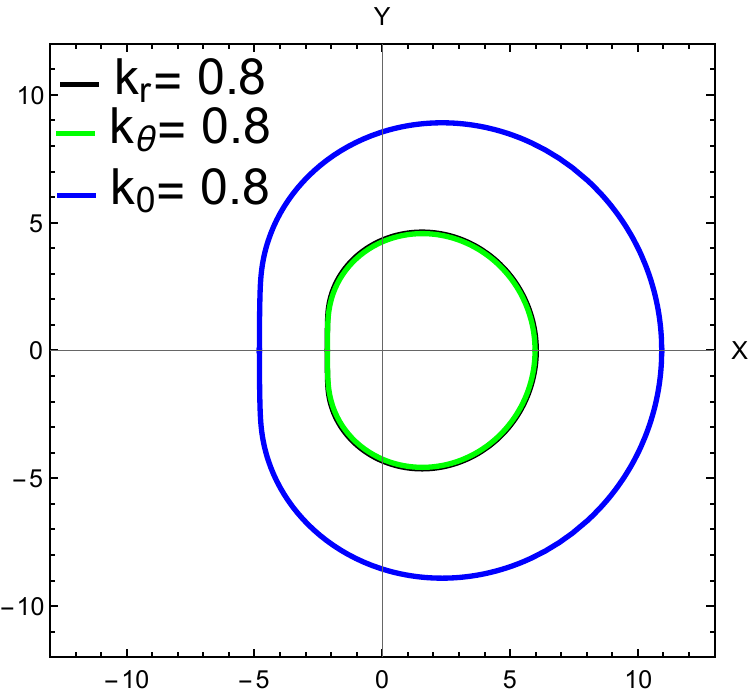}}}
     \\
     \caption{\label{figkkk}Shadow of a Kerr MOG black hole with homogeneous and nonhomogeneous plasma distributions. In the left plot, we take $a = 0.999$ with $\alpha = 0.0015$, whereas in the right plot, we take $a = 0.7$ with $\alpha = 1.04$. The shadow for the nonhomogeneous plasma environment appears smaller than the homogeneous one.}
     \end{figure}
      For the homogeneous ones
    {\begin{align}
         f_\text{r}(r)=k_0\omega_\text{{0}}^{2} r^{2}, \quad f_{\theta}(\theta)=k_0\omega_\text{{0}}^{2} a^{2} \cos ^{2} \theta
     \end{align}
    {while for the inhomogeneous case, we consider one of the distributions discussed in Sections B and C. The key finding in Fig. (\ref{figkkk}) is that the shadow in an inhomogeneous plasma environment is smaller than in a homogeneous one. We find that the shadow structures for the nonhomogeneous case almost overlap, with negligible deviation(right plot), which contrasts with the results obtained for the homogeneous case.
\section{Parameter Constraints Based on Black Hole Shadow }
\begin{justify}
The EHT Collaboration provided the first horizon-scale image of M87* in 2019 \cite{akiyama2019first}. With M87* at a distance of $d = 16.8 \, \text{Mpc}$ and an estimated mass of $M = (6.5 \pm 0.7) \times 10^9 M_\odot$ \cite{akiyama2019first}, the EHT results constrain the shadow angular diameter to $\theta_d = 42 \pm 3 \, \mu\text{as}$ and a circularity deviation $\Delta C \leq 0.1$. The shadow of M87* is consistent with $\mathrm{GTR}$  predictions for a Kerr BH; however, uncertainties in spin measurements and deviations in quadrupole moments allow for potential modifications \cite{akiyama2019first,cardoso2019testing} due to $\alpha$ and plasma effects. In 2022, EHT shadow observations of Sgr A* in the Milky Way revealed an angular diameter of $d_{\text{sh}} = 48.7 \pm 7 \, \mu\text{as}$ and a thick emission ring with $\theta_d = 51.8 \pm 2.3 \, \mu\text{as}$. With Sgr A* having a mass of $M = 4.0^{+1.1}_{-0.6} \times 10^6 M_\odot$ and a distance of $8 \, \text{kpc}$, these results align with predictions of $\mathrm{GTR}$ but also permit exploration of Kerr MOG BH scenarios.
Following the findings of \cite{akiyama2019event}, we use a distance to M87* of $D = (16.8 \pm 0.8)$ Mpc and a mass of $M = (6.5 \pm 0.7) \times 10^9 M_\odot$. From these values, the average diameter of the shadow can be expressed as \cite{akiyama2019event}
\[
\frac{d_{\text{sh}}}{M} = \frac{D  \theta_{\text{sh}}}{M} \approx 11.0 \pm 1.5,
\]
whereas the diameter of the shadow for Sgr A* is obtained as \cite{akiyama2022first}
\[
\frac{d_{\text{sh}}}{M} = \frac{D \theta_{\text{sh}}}{M} \approx 9.5 \pm 1.4.
\]
As we discussed above, rotating BHs produce shadows that differ significantly from those of nonrotating BHs, which are perfectly circular. When observed from non-polar directions, the shadow of a rotating BH appears displaced in the direction of rotation due to relativistic effects. For sufficiently high spin values, the shadow undergoes further distortion caused by the Lense-Thirring effect \cite{johannsen2013systematic,jafarzade2022observational}. Traditionally, the size and distortion of the shadow have been characterized by two parameters, $\delta_s$ and $R_s$, introduced by \cite{hioki2009measurement}. Here, $R_s$ represents the radius of a circle that approximates the shadow by passing through its top, bottom, and right edges, given as \cite{wei2013observing}
\begin{align}\label{rs}
    R_\text{s}=\frac{(X_\text{t}-X_\text{r})^2+ Y_\text{t}^2}{2(X_\text{r}-X_\text{t})}.
\end{align}
While $\delta_s$ quantifies the deviation of the shadow's left edge from this circular boundary \cite{hioki2009measurement} is $\delta_s=\frac{D_{cs}}{R_s}$,
where $D_{cs}$ is the difference between the left endpoints of the circle and left endpoints of the shadow \cite{abdujabbarov2015coordinate}. As a result, newer observables have been proposed \cite{schee2009optical, tsukamoto2014constraining, cunha2015shadows, younsi2016new, perlick2017light, wang2018chaotic}. 
The EHT has provided valuable constraints on BH parameters such as mass, but it cannot directly measure angular momentum \cite{akiyama2019first}. For instance, the EHT-derived mass of M87* aligns with estimates from stellar dynamics but deviates from measurements based on gas dynamics \cite{gebhardt2011black, walsh2013the}. To address the limitations of existing observables, new metrics for shadow characterization are proposed, including the shadow's area ($A$), circumference ($C$), and oblateness ($D$)  \cite{johannsen2011metric}. These observables allow for a general description of the shadow's geometry without relying on assumptions of circularity or symmetry, providing a more robust framework for studying BHs across diverse theoretical contexts \cite{kumar2020black}. 
The observables $A$ is defined by \cite{takahashi2004shapes,grenzebach2014photon}.
\begin{equation}\label{ad}
A = 2 \int{Y(r_\text{p}) dX(r_\text{p})} =\int_{r_{\text{p}}^{-}}^{r_{\text{p}}^{+}} \left(Y(r_\text{p} )\frac{dX(r_\text{p})}{r_\text{p}}\right)dr_\text{p}.
\end{equation}
 Observationally, it is feasible for an astronomer to measure the area, the boundary length, and the horizontal and vertical diameters of the shadow through precise astronomical imaging. These observables provide a unique characterization of the shadow, making it possible to estimate key BH parameters, such as spin and deviation from standard metrics, directly from observations.
 The recent EHT studies on the M87* BH observation have provided estimates of the Schwarzschild shadow deviation (\(\delta\)), a parameter quantifying the difference between the model shadow diameter (\(d_{\text{metric}}\)) and the Schwarzschild shadow diameter. This deviation is defined as \cite{akiyama2022first}
\begin{align}
\delta = \frac{\hat{d}_{\text{metric}}}{6\sqrt{3}} - 1, \quad \hat{d}_{\text{metric}} = 2R_a,
\end{align}
\begin{justify}
where \(R_a = \sqrt{\frac{A}{\pi}}\), with \(A\) being derived from (\ref{ad}). Notably, \(\delta\) can assume positive or negative values, depending on whether the observed shadow size is larger or smaller than that of a Schwarzschild BH of equivalent mass \cite{jafarzade2024kerr}. According to the results reported by the EHT observations, the measured Schwarzschild deviation is bounded as $\delta = -0.01 \pm 0.17$ \cite{akiyama2022first,akiyama2019first}.
\end{justify}
\begin{table}[h]
\centering
\caption{\label{tab1}Constraints on BH parameters using the EHT data of M87*.}
\resizebox{\textwidth}{!}{%
\begin{tabular}{|c|c|c|c|c|}
\hline
\multirow{3}{*}{Observable} & \multicolumn{2}{|c|}{Homogeneous plasma distribution} & \multicolumn{2}{|c|}{Inhomogeneous plasma distribution} \\ 
\cline{2-5}
& \multicolumn{2}{|c|}{$\alpha=0.2, k_0=0.5$} & \multicolumn{2}{|c|}{$\alpha=0.2, k_r=0.5$} \\ 
\cline{2-5}
& $1 \sigma$ & $2 \sigma$ & $1 \sigma$ & $2 \sigma$ \\ 
\hline
$d_{sh}$ & $a \in(0.76,0.9]$ & $a \in[0,0.76]$ & $a \in[0,0.75]$ & $a \in(0.75,0.9]$ \\ 
\hline
$\delta$ & -- & $a \in[0,0.9]$ & $a \in[0,0.9]$ & -- \\ 
\hline
\multirow{2}{*}{Observable} & \multicolumn{2}{|c|}{$a=0.5, k_0=0.5$} & \multicolumn{2}{|c|}{$a=0.5, k_r=0.5$} \\ 
\cline{2-5}
& $1 \sigma$ & $2 \sigma$ & $1 \sigma$ & $2 \sigma$ \\ 
\hline
$d_{sh}$ & $\alpha \in(0.39,3]$ & $\alpha \in[0,0.39]$ & $\alpha \in[0,0.41]$ & $\alpha \in(0.41,3]$ \\ 
\hline
$\delta$ & $\alpha \in[0.83,3]$ & $\alpha \in[0,0.83)$ & $\alpha \in[0,2.1]$ & $\alpha \in(2.1,3]$ \\ 
\hline
\multirow{2}{*}{Observable} & \multicolumn{2}{|c|}{$a=0.5, \alpha=0.2$} & \multicolumn{2}{|c|}{$a=0.5, \alpha=0.2$} \\ 
\cline{2-5}
& $1 \sigma$ & $2 \sigma$ & $1 \sigma$ & $2 \sigma$ \\ 
\hline
$d_{sh}$ & $k_0 \in[0,0.47)$ & $k_0 \in[0.47,0.603]$ & $k_r \in[0,1.23]$ & $k_r \in(1.23,5.15]$ \\ 
\hline
$\delta$ & $k_0 \in[0,0.417)$ & $k_0 \in[0.417,0.589]$ & $k_r \in[0,3.9]$ & $k_r \in(3.9,7.8]$ \\ 
\hline
\end{tabular}%
}
\label{tab:M87_constraints}
\end{table}

\begin{table}[h]
\centering
\caption{\label{tab2}Constraints on BH parameters using the EHT data of Sgr A*.}
\resizebox{\textwidth}{!}{%
\begin{tabular}{|c|c|c|c|c|}
\hline
\multirow{3}{*}{Observable} & \multicolumn{2}{|c|}{Homogeneous plasma distribution} & \multicolumn{2}{|c|}{Inhomogeneous plasma distribution} \\ 
\cline{2-5}
& \multicolumn{2}{|c|}{$\alpha=0.1, k_0=0.2$} & \multicolumn{2}{|c|}{$\alpha=0.1, k_r=0.2$} \\ 
\cline{2-5}
& $1 \sigma$ & $2 \sigma$ & $1 \sigma$ & $2 \sigma$ \\ 
\hline
$d_{sh}$ & $a \in(0.48,0.9]$ & $a \in[0,0.48]$ & $a \in[0,0.9]$ & -- \\ 
\hline
$\delta$ (VLTI) & -- & $a \in[0,0.9]$ & $a \in[0,0.9]$ & -- \\ 
\hline
$\delta$ (Keck) & $a \in(0.52,0.9]$ & $a \in[0,0.52]$ & $a \in[0,0.9]$ & -- \\ 
\hline
\multirow{2}{*}{Observable} & \multicolumn{2}{|c|}{$a=0.5, k_0=0.2$} & \multicolumn{2}{|c|}{$a=0.5, k_r=0.2$} \\ 
\cline{2-5}
& $1 \sigma$ & $2 \sigma$ & $1 \sigma$ & $2 \sigma$ \\ 
\hline
$d_{sh}$ & $\alpha \in(0.092,3]$ & $\alpha \in[0,0.092]$ & $\alpha \in[0,3]$ & -- \\ 
\hline
$\delta$ (VLTI) & $\alpha \in(0.041,3]$ & $\alpha \in[0,0.041]$ & $\alpha \in[0,2.25]$ & $\alpha \in(2.25,3]$ \\ 
\hline
$\delta$ (Keck) & $\alpha \in(0.108,3]$ & $\alpha \in[0,0.108]$ & $\alpha \in[0,1.5]$ & $\alpha \in(1.5,3]$ \\ 
\hline
\multirow{2}{*}{Observable} & \multicolumn{2}{|c|}{$a=0.5, \alpha=0.1$} & \multicolumn{2}{|c|}{$a=0.5, \alpha=0.1$} \\ 
\cline{2-5}
& $1 \sigma$ & $2 \sigma$ & $1 \sigma$ & $2 \sigma$ \\ 
\hline
$d_{sh}$ & $k_0 \in[0,0.203]$ & $k_0 \in(0.203,0.427]$ & $k_r \in[0,5.6)$ & $k_r \in[5.6,9.5]$ \\ 
\hline
$\delta$ (VLTI) & $k_0 \in[0,0.103]$ & $k_0 \in(0.103,0.297]$ & $k_r \in[0,4.1]$ & $k_r \in(4.1,6.4]$ \\ 
\hline
$\delta$ (Keck) & $k_0 \in[0,0.197]$ & $k_0 \in(0.197,0.361]$ & $k_r \in(0,3.4)$ & $k_r \in[3.4,5.9]$ \\ 
\hline
\end{tabular}%
}
\label{tab:SgrA_constraints}
\end{table}
The EHT utilized two distinct priors for the angular size of Sgr A*, derived from observations by the Keck and the Very Large Telescope Interferometer (VLTI), to estimate the bounds on the fractional deviation observable $\delta$, given as \cite{akiyama2022first}.
\[
\delta_{\text{Sgr}} = 
\begin{cases} 
-0.08^{+0.09}_{-0.09} & \text{(VLTI)} \\
-0.04^{+0.09}_{-0.10} & \text{(Keck)} 
\end{cases}
\]
\begin{justify}
Calculating the mentioned observables, we can constrain the plasma and other parameters by the EHT for the supermassive BH M87* and Sgr A*. Table \ref{tab1} shows the allowed regions of parameters that satisfy the mentioned constraint of M87* data. As we see, in a homogeneous plasma spacetime and for a small deformation parameter $\alpha$, the resulting shadow is consistent with EHT data within $ 2\sigma $ uncertainty for the range {$0 < a < 0.76$}. While in a nonhomogeneous plasma spacetime, the resulting shadow is located in $ 1\sigma $ confidence region for the mentioned range. In another analysis, we fixed the plasma and rotation parameters and investigated the allowed range of the deformation parameter.

We noticed that in a homogeneous (nonhomogeneous) plasma spacetime and for intermediate values of the rotation parameter, the resulting shadow is in agreement with observations data of M87* within $ 1\sigma $ ($ 2\sigma $) uncertainty for the wide range of $ \alpha $. We also fixed the rotation and deformation parameters and found the allowed range of the plasma parameter. According to our findings, the probability of finding results consistent with observational data is higher in a nonhomogeneous plasma distribution than in a homogeneous distribution.

We continue our analysis by constraining parameters using EHT data of Sgr A*. The allowed range of parameters is addressed in Table \ref{tab2}. It can be seen that the resulting shadow is more consistent with observational data in a nonhomogeneous environment. This table also indicates the allowable region of parameters that satisfy the Keck and VLTI bounds. In a nonhomogeneous plasma distribution, all values of the rotation parameter satisfy both VLTI and Keck bounds within $ 1\sigma $ uncertainty. While in a homogeneous plasma distribution, no values of $a$ satisfy $ 1\sigma $ VLTI bound and only the range $0.52<a<0.9$ satisfies $ 1\sigma $ Keck bound. For the fixed rotation and plasma parameters and in both homogeneous/nonhomogeneous plasma distributions, a range of $ \alpha $ which satisfies $ 1\sigma $ VLTI bound is wider than a range that satisfies $ 1\sigma $ Keck bound. Comparing Table \ref{tab1} to Table \ref{tab2}, one can find that for small plasma parameters, Sgr A* BH can be a suitable model for MOG BHs, while for intermediate values of $ k_{0}/k_{r} $, M87* BH is a suitable model for MOG BHs. 

To find the allowed region of parameters in latitudinal plasma distribution, we plotted Figs. \ref{Figeht1}-\ref{Figeht3}. From Fig. \ref{Figeht1}, it can be seen that the allowed range of parameters $a$, $\alpha$ and $k_{\theta}$ for which $d_{sh}$ agrees with the Sgr A* data is wider than the range that matches the M87* data. Fig. \ref{Figeht2} shows the regions of parameters that satisfy the constraint on the deviation $ \delta $ using values reported by M87* measurements. 
 As already mentioned, $ \delta $ can be negative (positive) if the BH shadow size is smaller (greater) than the Schwarzschild BH of the same mass. Looking at Fig. \ref{Figeht2}, one notices that for a small (large) latitudinal plasma
parameter, a MOG BH with the same mass as Schwarzschild, has a smaller (greater) shadow size compared to the Schwarzschild BH. To estimate upper and lower bounds
of parameters using VLTI and Keck measurements, we have plotted Fig. \ref{Figeht3}. It is clear that the ranges of parameters satisfying the VLTI bound are wider than those of the Keck bound. 
\end{justify}
\begin{figure}[!htb]
\centering
\subfloat[$ \alpha=0.2 $]{
        \includegraphics[width=0.49\textwidth]{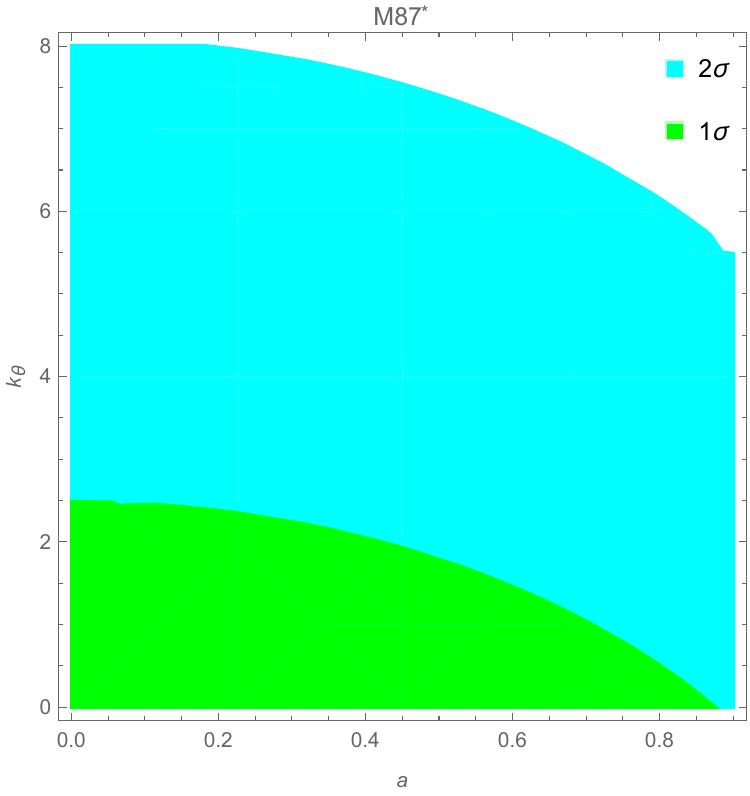}}
\subfloat[$ a=0.5 $]{
        \includegraphics[width=0.49\textwidth]{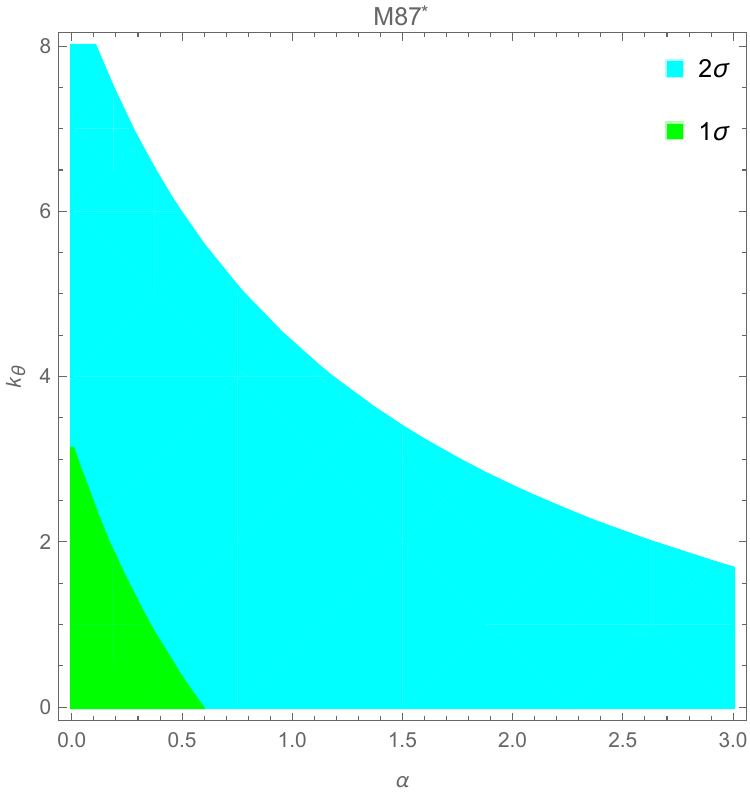}}
\newline
\subfloat[$ \alpha=0.2 $]{
        \includegraphics[width=0.49\textwidth]{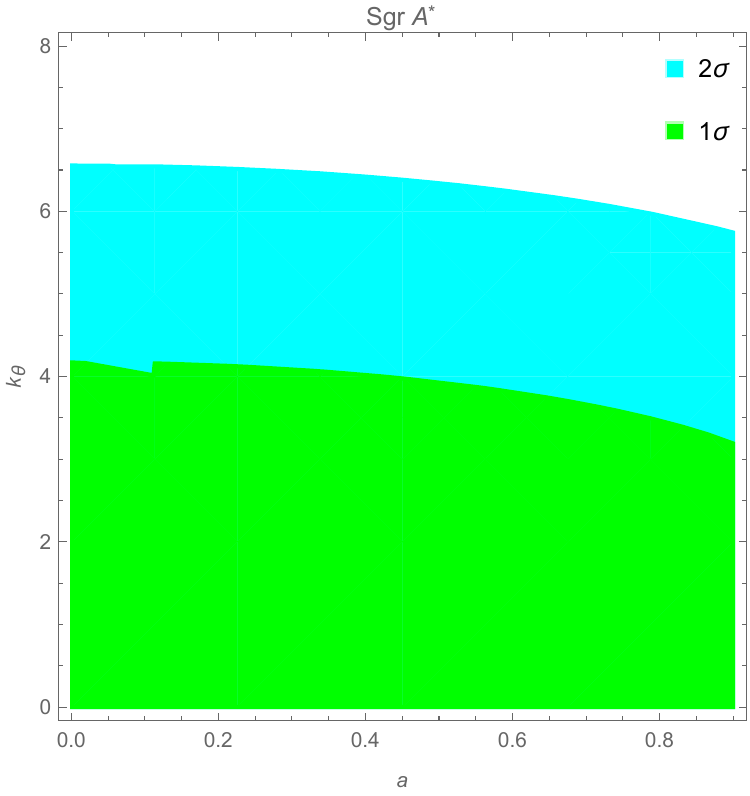}}
\subfloat[$ a=0.5 $]{
        \includegraphics[width=0.49\textwidth]{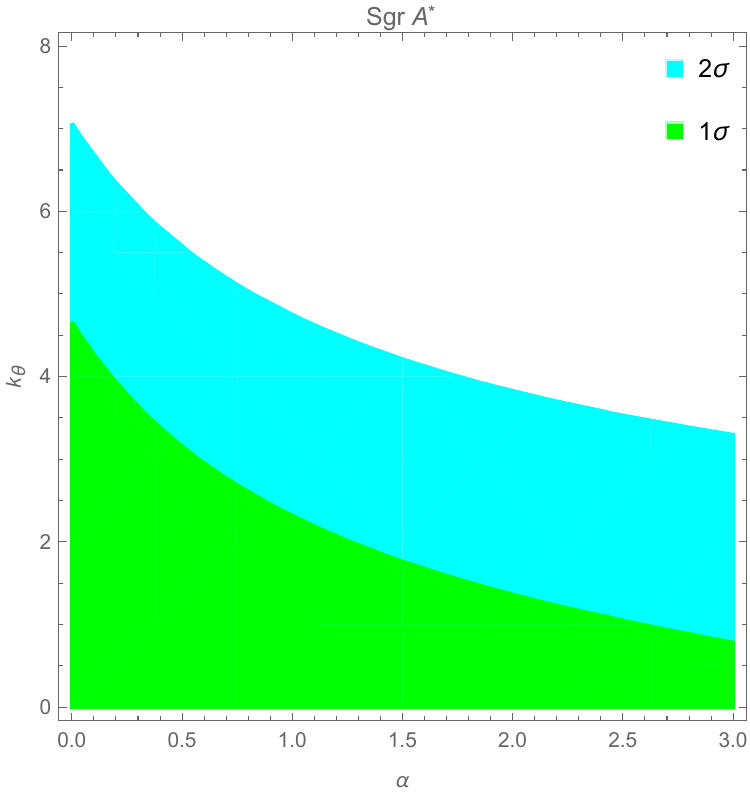}}        
\newline
\caption{Constraints on the BH parameters with the EHT observations
of M87* and Sgr A*.}
\label{Figeht1}
\end{figure}
\begin{figure}[!htb]
\centering
\subfloat[$ \alpha=0.2 $]{
        \includegraphics[width=0.45\textwidth]{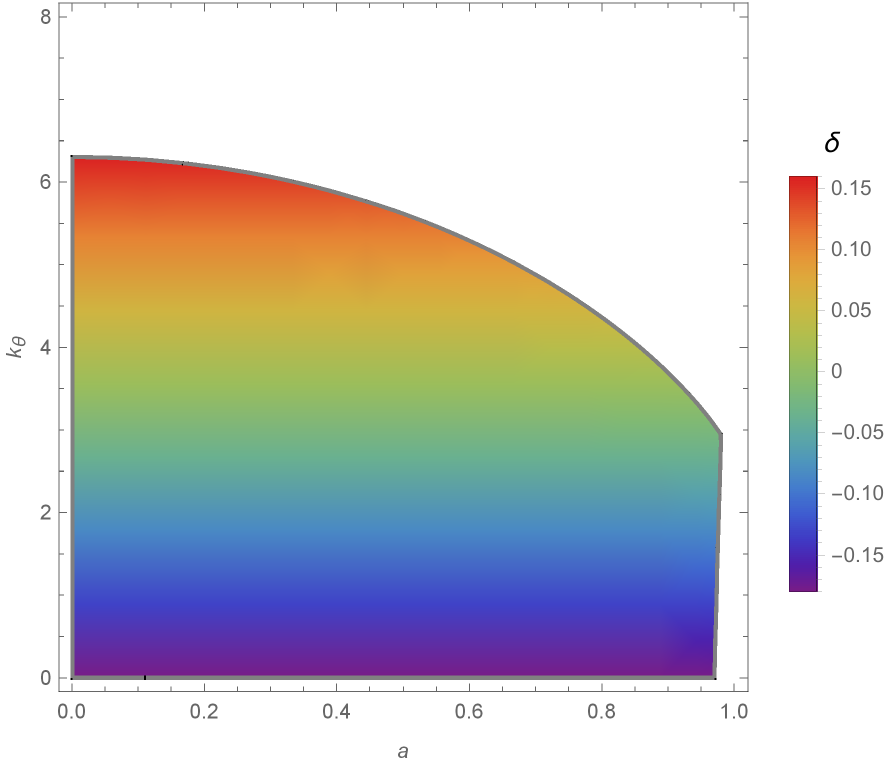}}
\subfloat[$ a=0.5 $]{
        \includegraphics[width=0.45\textwidth]{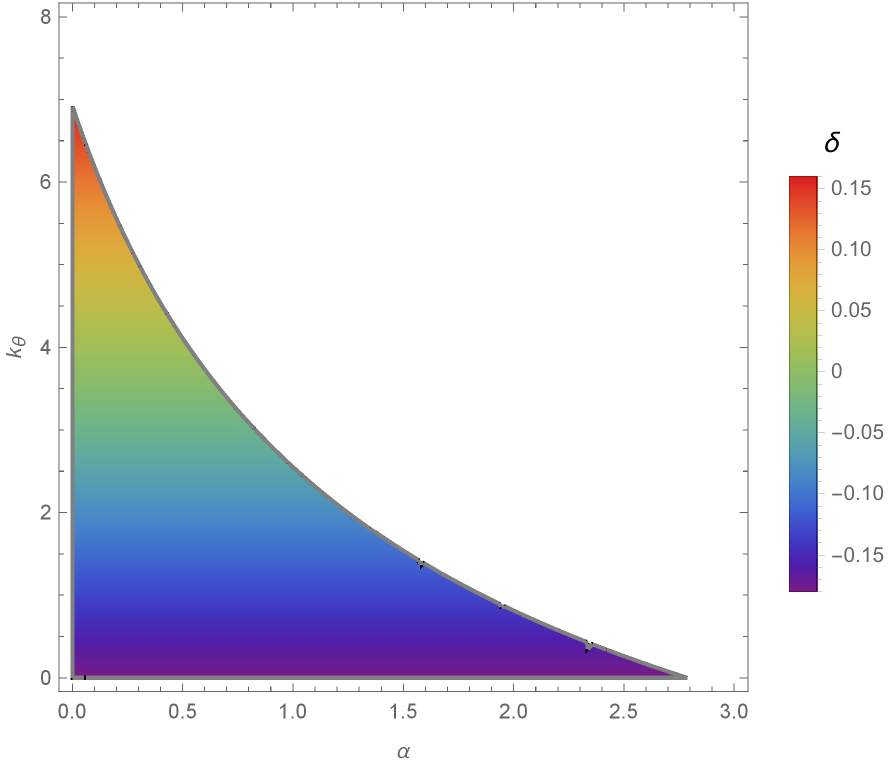}}       
\newline
\caption{The shadow diameter deviation from that of a Schwarzschild BH as a function of $a$, $\alpha$, and $k_{\theta}$ based on M87* data.}
\label{Figeht2}
\end{figure}

\begin{figure}[!htb]
\centering
\subfloat[$ \alpha=0.2 $]{
        \includegraphics[width=0.45\textwidth]{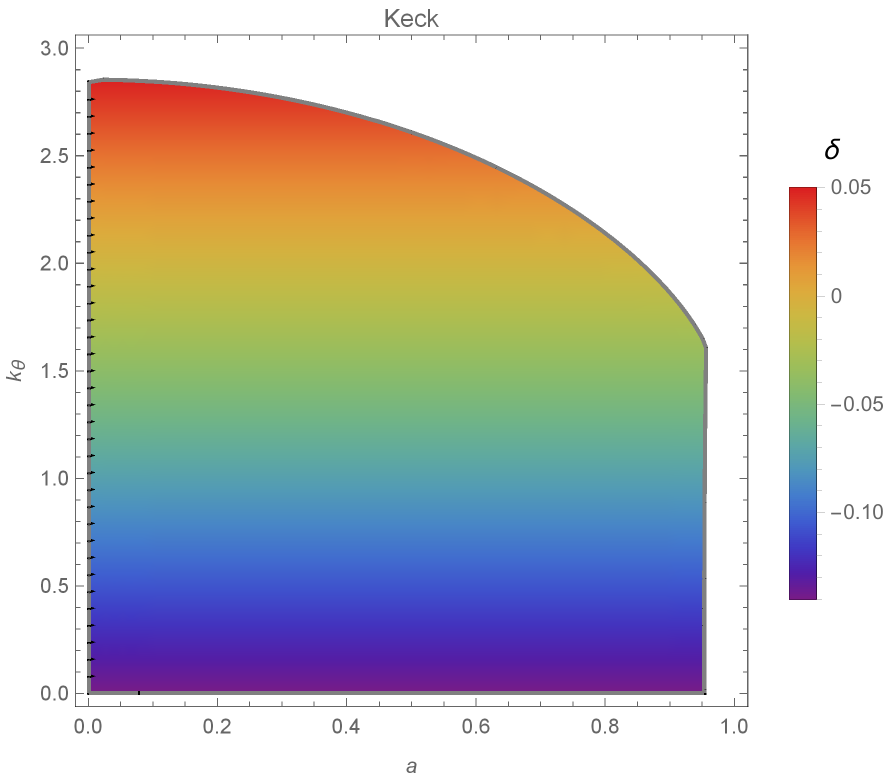}}
\subfloat[$ a=0.5 $]{
        \includegraphics[width=0.45\textwidth]{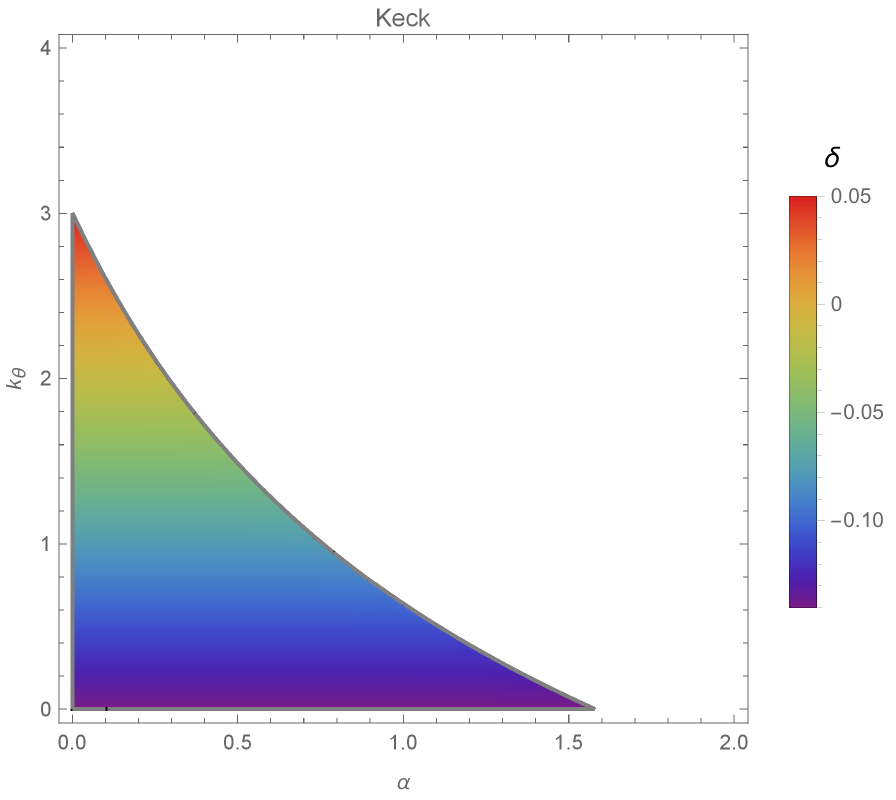}}
\newline
\subfloat[$ \alpha=0.2 $]{
        \includegraphics[width=0.45\textwidth]{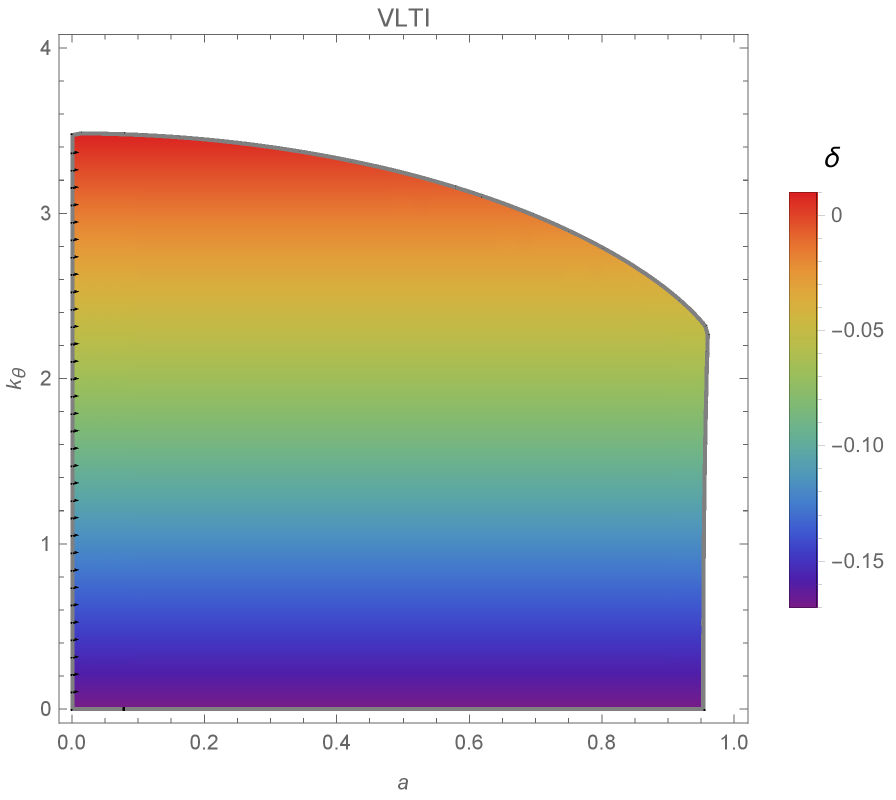}}
\subfloat[$ a=0.5 $]{
        \includegraphics[width=0.45\textwidth]{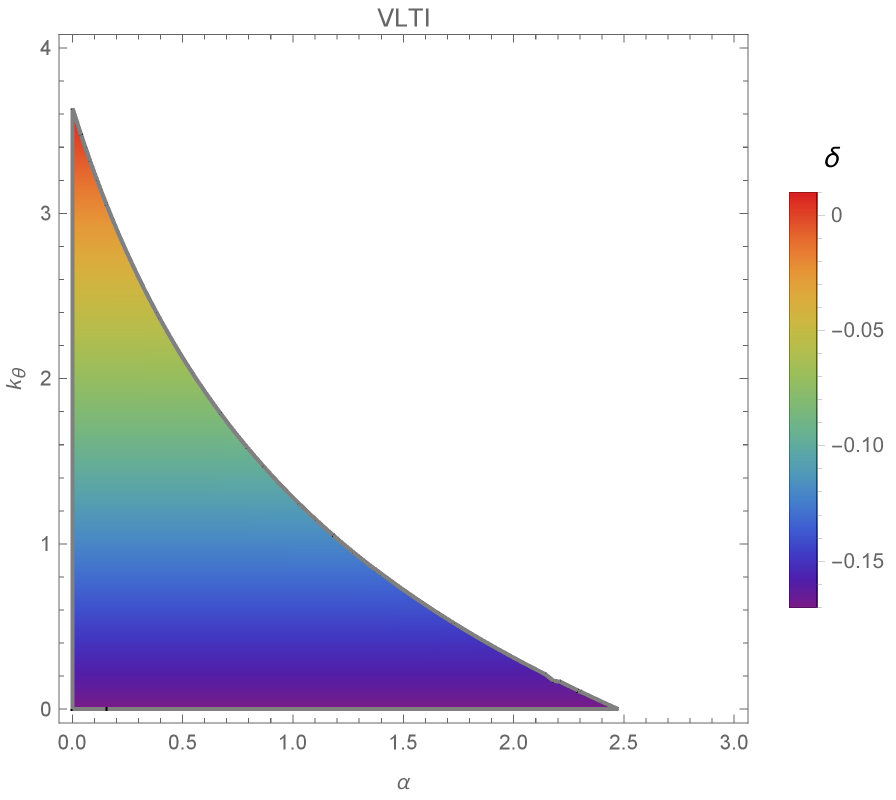}}        
\newline
\caption{The shadow diameter deviation from that of a Schwarzschild BH as a function of $a$, $\alpha$, and $k_{\theta}$ based on Sgr A* data.}
\label{Figeht3}
\end{figure}
\end{justify}
\section{Energy Emission Rate}
The energy emission rate associated with BHs, particularly through the mechanism known as Hawking radiation, is a fascinating interplay of quantum mechanics and $\mathrm{GTR}$. This process involves the creation of virtual particle pairs near the event horizon
of a BH, where one particle escapes while the other falls into the BH, effectively reducing its mass \cite{jafarzade2024thermodynamics}.
For Kerr MOG BHs, this analysis becomes more complex due to the modifications in spacetime geometry introduced by the STVG framework of the theory. These modifications influence critical parameters such as the Hawking temperature. By extending the methods of analyzing Hawking radiation to the Kerr MOG scenario, it is possible to explore how modified gravity affects BH thermodynamics and energy emission characteristics. According to \cite{wei2013observing}, the energy emission rate can be expressed in its general form as
\begin{align}   
\frac{d^2E(\omega)}{d\omega dt} = \frac{2\pi^3 R_\text{s}^2\omega^3}{e^{\omega / T_\text{H}} - 1},
\end{align}
where the $\omega$ is frequency of photons and $R_\text{s}$ radius of the shadow given in (\ref{rs}). The  corresponding formula of the Hawking temperature reads as
\begin{align}
    T_\text{H}= \frac{2\left(r(1+\alpha)-\mathcal{M}\right)}{4\pi(1+\alpha)(r_{e}^2+a^2)},
\end{align}
where $r_\text{e}$ is the event horizon of the BH. To show how the energy emission rate is affected by the parameters of the model, we have plotted Fig. \ref{FigEr}.  Fig. \ref{FigEr}(a) illustrates the influence of the homogeneous plasma parameter $ k_{0} $ on the emission rate, verifying that the evaporation process would be faster for a
BH is located in a homogeneous plasma spacetime. In other words, the BH has a shorter lifetime in a homogeneous plasma distribution. From Figs. \ref{FigEr}(b) and \ref{FigEr}(c), it can be seen that the effect of nonhomogeneous plasma parameters on the emission rate is opposite to $ k_{0} $ effect, meaning that the BH has a longer lifetime in a nonhomogeneous plasma spacetime. Comparing the figures \ref{FigEr}(b), \ref{FigEr}(c) with \ref{FigEr}(a), one can notice that nonhomogeneous plasma parameters have an insignificant effect on the emission rate in comparison with the homogeneous plasma parameter. Regarding the effect of the rotation parameter $ a $ and deformation parameter $ \alpha $ on this optical quantity, Figs. \ref{FigEr}(d) and \ref{FigEr}(e) depicts both parameters decrease the emission rate, meaning that the evaporation process would be slower for fast-rotating BHs.
 \section{Deflection Angle}
       \begin{justify}
       In this section, we investigate the influence of plasma
distributions on the deflection angle within the deformation parameter. When the value of $\frac {\omega_\text{p}^2}{\omega_0^2}$ increases from critical value, a photon arriving from infinity will be reflected back to infinity after reaching a certain distance. To a distant observer, the photon's trajectory will not appear as a straight line but will be bent due to the curvature of spacetime.
\begin{justify}
       Here, we aim to examine the deflection angle of a photon as it passes by the BH. From the equations of motion, we get the following 
\end{justify}
\begin{align}\label{hrr}
\frac{dr}{d\phi} = \frac{\dot{r}}{\dot{\phi}} = \frac{g^{rr} p_r}{g^{\phi\phi} p_\phi + g^{t\phi} p_t}.
\end{align}
As $-p_t= \omega_0$ and considering by  $H=0$, we can rewrite above expression as \cite{wang2022shadow}
\begin{figure}[H]
    \centering
    \begin{subfigure}{0.45\textwidth}
        \centering
        \includegraphics[height=4cm,width=7cm]{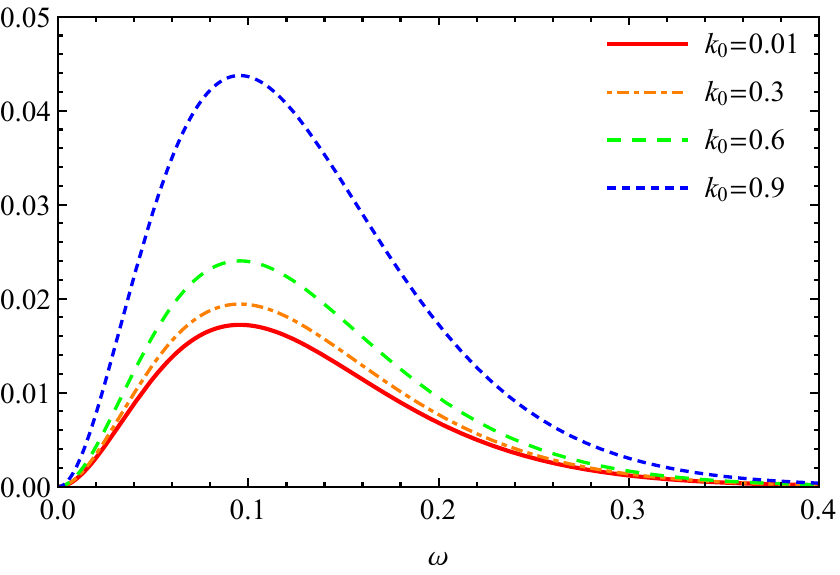}
        \caption{$a=0.3$ and $\alpha=0.1$}
    \end{subfigure}
    \hfill
    \begin{subfigure}{0.45\textwidth}
        \centering
        \includegraphics[height=4cm,width=7cm]{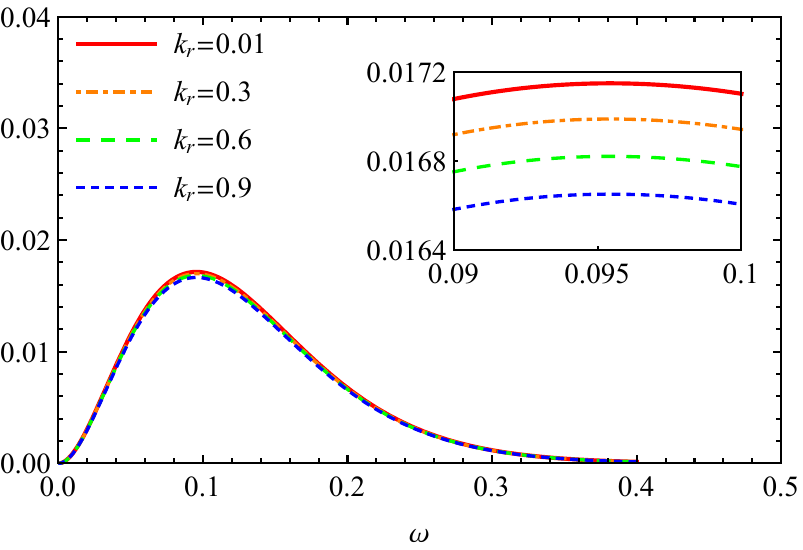}
        \caption{$a=0.3$ and $\alpha=0.1$}
    \end{subfigure}
    
    \vspace{0.5cm}  

    \begin{subfigure}{0.45\textwidth}
        \centering
        \includegraphics[height=4cm,width=7cm]{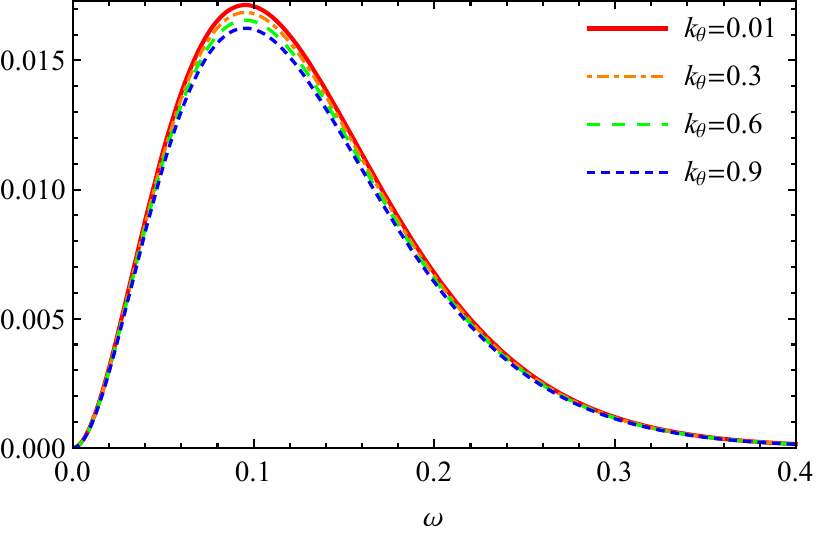}
        \caption{$a=0.3$ and $\alpha=0.1$}
    \end{subfigure}
    \hfill
    \begin{subfigure}{0.45\textwidth}
        \centering
        \includegraphics[height=4cm,width=7cm]{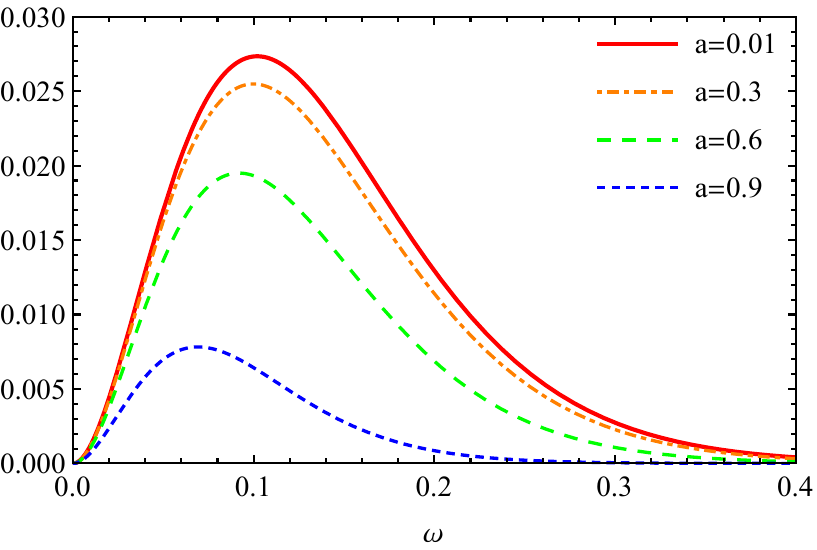}
        \caption{ $\alpha=0.1$ and $k_0=0.5$}
    \end{subfigure}

    \vspace{0.5cm}

    \begin{subfigure}{0.45\textwidth}
        \centering
        \includegraphics[height=5cm,width=7cm]{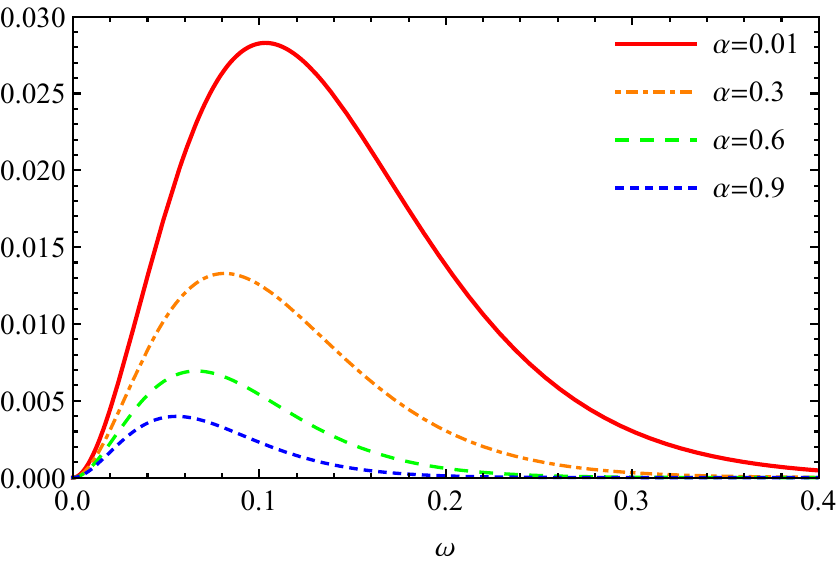}
        \caption{$a=0.1$ and $k_0=0.5$}
    \end{subfigure}
    
    \caption{\label{FigEr} Energy emission rates \( \frac{d^2E(\omega)}{d\omega dt} \) is plotted on the vertical axis against frequency \( \omega \) for different values of  \( a \), \( \alpha \), \( k_0 \), \( k_r \), and \( k_{\theta} \).}
\end{figure}
\end{justify}
where 
\begin{justify}
\begin{equation}
h(r)^{2}=-\frac{g^{t t}}{g^{\phi \phi}}+\frac{g^{t \phi} g^{t \phi}}{g^{\phi \phi} g^{\phi \phi}}-\frac{k}{g^{\phi \phi}} ,
\end{equation}
where $k=(k_0,k_r,k_\theta)$. At the minimum distance $R, d r / d \phi=0$ is satisfied, and therefore we have
\begin{equation}
h(R)^{2}=\frac{\left(p_{\phi}-\frac{g^{t \phi}}{g^{\phi \phi}} \omega_{0}\right)^{2}}{\omega_{0}^{2}} .
\end{equation}

\end{justify}
\begin{figure}[H]
    \centering
    \begin{subfigure}{0.45\textwidth}
        \centering
        \includegraphics[height=4cm,width=7cm]{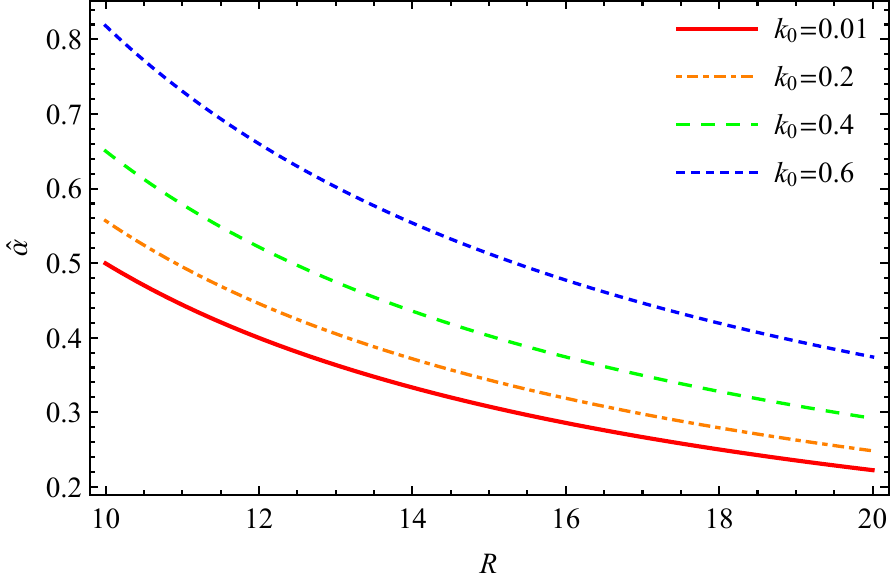}
        \caption{ $a=0.3$ and $\alpha=0.1$}
    \end{subfigure}
    \hfill
    \begin{subfigure}{0.45\textwidth}
        \centering
        \includegraphics[height=4cm,width=7cm]{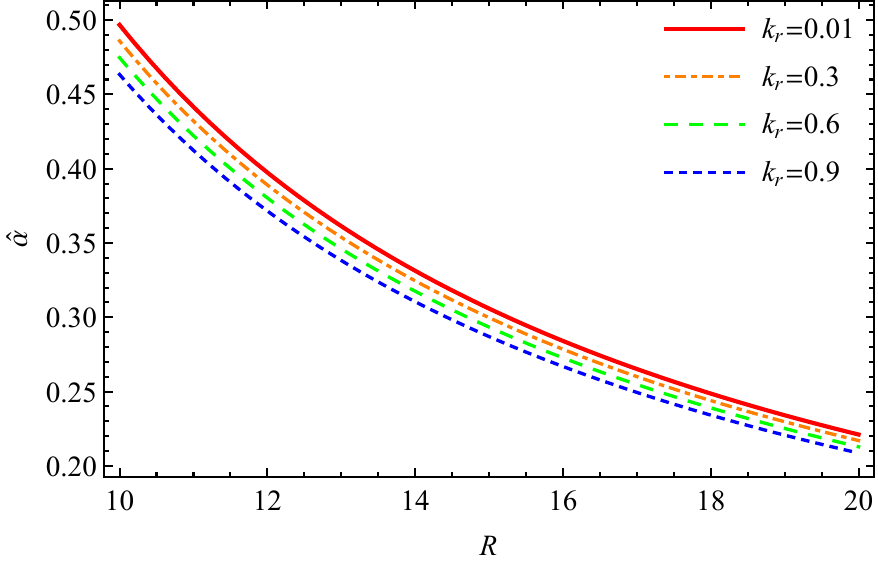}
        \caption{$a=0.3$ and $\alpha=0.1$}
    \end{subfigure}
     \vspace{0.5cm}
    \begin{subfigure}{0.45\textwidth}
        \centering
        \includegraphics[height=4cm,width=7cm]{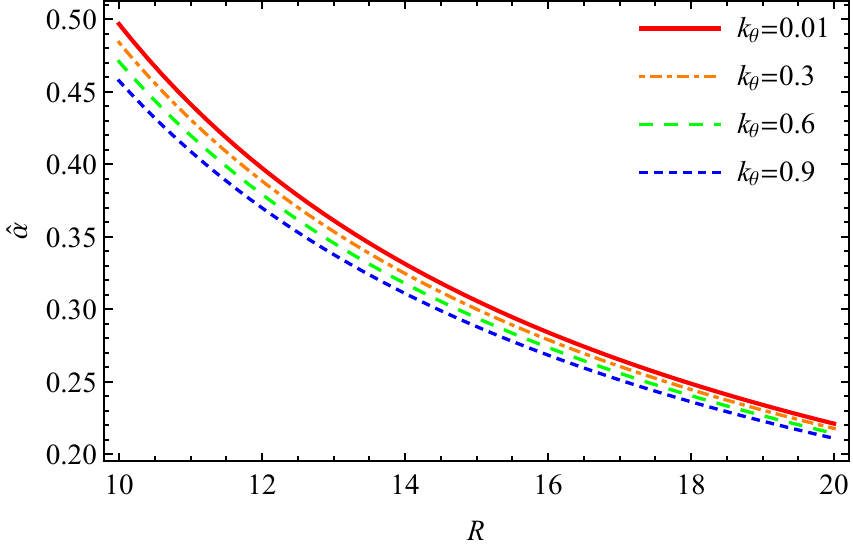}S
        \caption{$a=0.3$ and $\alpha=0.1$}
    \end{subfigure}
    \hfill
    \begin{subfigure}{0.45\textwidth}
        \centering
        \includegraphics[height=4cm,width=7cm]{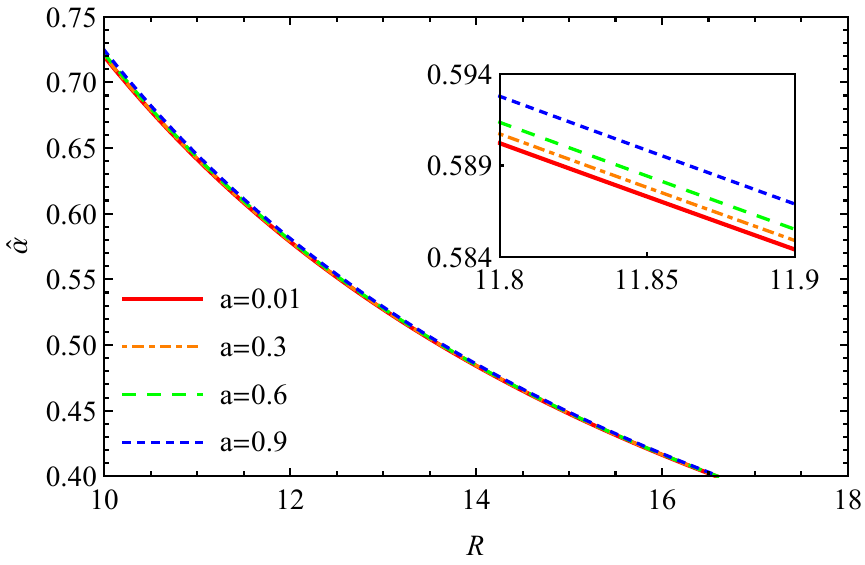}
        \caption{ $\alpha=0.1$ and $k_0=0.5$}
    \end{subfigure}
     \vspace{0.5cm}
    \begin{subfigure}{0.45\textwidth}
        \centering
        \includegraphics[height=5cm,width=7cm]{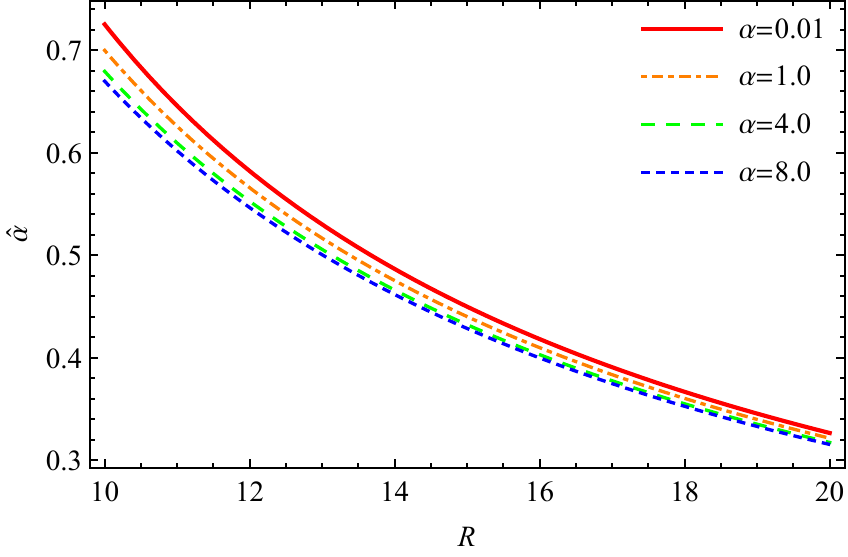}
        \caption{$a=0.3$ and $k_0=0.5$}
    \end{subfigure}
    
    \caption{\label{Figdef} Deflection angle vs distance $R$ for different values of $ a $, $ \alpha $, $ k_{0}$, $ k_{r} $, and $ k_{\theta} $.}
\end{figure}
Integrating (\ref{hrr}) with respect to $r$, we get the representation of the deflection angle of the light ray in the Kerr MOG spacetime given by \cite{wang2022shadow}
\begin{align}
\hat{\alpha} & =2 \int_{R}^{\infty} \sqrt{\frac{g^{\phi \phi}}{g^{r r}}}\left(\frac{h(r)^{2}}{h(R)^{2}}-1\right)^{-1 / 2} d r-\pi. 
\end{align}
\begin{justify}
The behavior of the deflection angle of light  $ \hat{\alpha} $  concerning the impact parameter $ b $ is illustrated in Fig. \ref{Figdef}. From Fig. \ref{Figdef}(a), one can find that light rays are more deflected by BHs in a homogeneous plasma spacetime, while an opposite behavior is observed in a nonhomogeneous plasma spacetime (see Figs. \ref{Figdef}(b) and \ref{Figdef}(c)). Fig. \ref{Figdef}(d) displays the effect of the spin parameter on $ \hat{\alpha} $ and confirms that increasing this parameter leads to an increase in the deflection angle. The influence of the deformation parameter on $ \hat{\alpha} $ is illustrated in Fig. \ref{Figdef}(e), verifying that the deflection angle decreases as $\alpha$ increases.
\end{justify}
\section{Conclusion}
\begin{justify}
We investigated light propagation in a non-magnetized, pressureless plasma which is a dispersive medium with a frequency-dependent refractive index on Kerr MOG spacetime. The gravitational field is determined solely by the mass, spin, and MOG parameter $\alpha$ of the Kerr MOG BH thus, the gravitational field of the plasma particles is not included in this analysis. In this model, the presence of the plasma affects only the trajectories of the light rays.
 We applied the Hamiltonian formalism to describe photon dynamics within the plasma environment. Here, we utilized the $\mathrm{HJ}$ equations and separated the equation by applying a specific condition on the plasma density and attained the generalized Carter constant \cite{perlick2017light}. Based on the separability condition, we identify photon regions that meet this criterion and provide an analytical formula for the boundary curve of the shadow. 

It was observed that the spacetime model influenced the shape and size of the shadow in distinct ways, with the deformation parameter \( \alpha \) playing a crucial role.
If the plasma frequency is much smaller than the photon frequency, the shadow closely resembles the pure gravity case. However, when the plasma frequency approaches the photon frequency, the shadow’s properties change significantly depending on the plasma distribution. Notably, there exists a certain plasma frequency ratio, \( \frac{\omega_p^{2}}{\omega_0^{2}} \), above which the shadow becomes invisible. This threshold depends on the metric and the \( \alpha \) parameter. Consequently, the size and shape of the shadow are altered in the presence of a plasma environment around the BH, depending significantly on the ratio between the plasma frequency and photon frequency, as well as the deformation parameter \( \alpha \). Then our analysis revealed that the shadow characteristics of M87* and Sgr A* are more consistent with EHT observational data in nonhomogeneous plasma spacetime compared to homogeneous distributions. For small deformation parameter $\alpha$, the shadow aligns within $2\sigma$ uncertainty in a homogeneous plasma and within $1\sigma$ in a nonhomogeneous plasma.

The probability of finding the results  
consistent with observational data is higher in a nonhomogeneous plasma distribution than in a homogeneous distribution. These findings underscore the role of plasma properties and spacetime deformations in refining models of supermassive BHs. Using EHT data of Sgr A* highlights that the resulting shadow is more consistent with observational data in a nonhomogeneous plasma environment. In a nonhomogeneous plasma distribution, all values of the rotation parameter $a$ satisfy both VLTI and Keck bounds within $1\sigma$ uncertainty, while in a homogeneous plasma distribution, no value of $a$ satisfies the $1\sigma$ VLTI bound, and only a limited range satisfies the $1\sigma$ Keck bound. For fixed rotation and plasma parameters, the range of the deformation parameter $\alpha$ satisfying the $1\sigma$ VLTI bound is broader than the range satisfying the $1\sigma$ Keck bound. Comparing Sgr A* and M87*, we find that for a small plasma parameter, Sgr A* serves as a more suitable model for MOG BHs, while for intermediate values of $k_{0}$ and $k_{r}$, M87* align better. 

 The energy emission rate of BHs is significantly affected by the plasma distribution and the model parameter $\alpha$. The BH undergoes a faster evaporation process in a homogeneous plasma spacetime, leading to a shorter lifetime. In contrast, nonhomogeneous plasma parameters extend the lifetime of BH, having an opposite effect on the emission rate compared to the homogeneous plasma parameter. Furthermore, the impact of nonhomogeneous plasma parameters on the emission rate is relatively minor compared to that of the homogeneous plasma parameter. The spin parameter $a$ and the deformation parameter $\alpha$ also decrease the emission rate, indicating that fast-rotating BHs and those with larger deformation parameters evaporate more slowly, resulting in longer lifetimes. Finally, we found that light rays are more strongly deflected by BHs in a homogeneous plasma spacetime, while the deflection angle is reduced in a nonhomogeneous plasma spacetime. The spin parameter $a$ enhances the deflection angle, indicating that BHs with higher rotation parameters bend light more significantly. In contrast,  $\alpha$ decreases the deflection angle, suggesting that greater deformation reduces the gravitational lensing effect of  BH.
  \end{justify}
\bibliographystyle{ieeetr}
\bibliography{main}

\begin{thebibliography}{10}

\bibitem{zwicky1933redshift}
F.~Zwicky {\em et~al.}, ``The redshift of extragalactic nebulae,'' {\em Helv. Phys. Acta}, vol.~6, 138(1933).

\bibitem{rubin1980rotational}
V.~C. Rubin, N.~Thonnard, and W.~K.~J. Ford, ``Rotational properties of 21 {SC} galaxies with a large range of luminosities and radii, from {NGC} 4605$(r= 4 kpc)$ to {UGC} 2885 $(r= 122 kpc)$,'' {\em Astrophysical Journal}, vol.~238, 487(1980).

\bibitem{lee2017innermost}
H.~C. Lee and Y.~J. Han, ``Innermost stable circular orbit of {K}err-{MOG} black hole,'' {\em The European Physical Journal C}, vol.~77, 9(2017).

\bibitem{milgrom1983modification}
M.~Milgrom, ``A modification of the newtonian dynamics as a possible alternative to the hidden mass hypothesis,'' {\em Astrophysical Journal}, vol.~270, 370(1983).

\bibitem{bhattacharjee2014rotation}
P.~Bhattacharjee, S.~Chaudhury, and S.~Kundu, ``Rotation {C}urve of the {M}ilky {W}ay out to~ 200 kpc,'' {\em The Astrophysical Journal}, vol.~785, 63(2014).

\bibitem{moffat2008testing}
J.~Moffat and V.~T. Toth, ``Testing modified gravity with globular cluster velocity dispersions,'' {\em The Astrophysical Journal}, vol.~680, 1158(2008).

\bibitem{moffat2006scalar}
J.~W. Moffat, ``Scalar {T}ensor {V}ector {G}ravity {T}heory,'' {\em Journal of Cosmology and Astroparticle Physics}, vol.~2006, 004(2006).

\bibitem{brownstein2007bullet}
J.~R. Brownstein and J.~W. Moffat, ``The {B}ullet {C}luster 1{E}0657-558 evidence shows modified gravity in the absence of dark matter,'' {\em Monthly Notices of the Royal Astronomical Society}, vol.~382, 47(2007).

\bibitem{moffat2015black}
J.~W. Moffat, ``Black holes in modified gravity {(MOG)},'' {\em The European Physical Journal C}, vol.~75, 175(2015).

\bibitem{Abbott2016observation}
B.~P. Abbott, R.~Abbott, T.~D. Abbott, {\em et~al.}, ``Observation of gravitational waves from a binary black hole merger,'' {\em Physical Review D}, vol.~116, 061102(2016).

\bibitem{raidal2017observation}
M.~Raidal, V.~Vaskonen, and H.~Veermäe, ``Observation of gravitational waves from a binary black hole merger,'' {\em Journal of Cosmology and Astroparticle Physics}, vol.~9, 37(2017).

\bibitem{cardenas2016testing}
A.~Cárdenas-Avendaño, J.~Jiang, and C.~Bambi, ``Testing the {K}err black hole hypothesis: comparison between the gravitational wave and the iron line approaches,'' {\em Physics Letters B}, vol.~760, 258(2016).

\bibitem{jafarzade2023rotating}
K.~Jafarzade, E.~Rezaei, and S.~Hendi, ``Rotating lifshitz-like black holes in {F} {(R)} gravity,'' {\em Progress of Theoretical and Experimental Physics}, vol.~2023, no.~5, 053E01(2023).

\bibitem{akiyama2019first}
K.~Akiyama, A.~Alberdi, {\em et~al.}, ``First {M87*} {E}vent {H}orizon {T}elescope {R}esults. {I}. the shadow of the supermassive black hole,'' {\em The Astrophysical Journal Letters}, vol.~875, 6(2019).

\bibitem{hendi2023blackhole}
S.~Hendi, K.~Jafarzade, and B.~E. Panah, ``Black holes in drgt massive gravity with the signature of eht observations of m87,'' {\em Journal of Cosmology and Astroparticle Physics}, vol.~2023, no.~02, p.~022, 2023.

\bibitem{jafarzade2024study}
K.~Jafarzade, Z.~Bazyar, and M.~Jamil, ``A study of black holes in {F} ({R})-{M}odmax gravity: Shadow and gravitational lensing,'' {\em arXiv preprint arXiv:2411.15757}, 2024.

\bibitem{jafarzade2021shadow}
K.~Jafarzade, M.~K. Zangeneh, and F.~S. Lobo, ``Shadow, deflection angle and quasinormal modes of {B}orn-{I}nfeld charged black holes,'' {\em Journal of Cosmology and Astroparticle Physics}, vol.~2021, 008(2021).

\bibitem{Synge1966thescape}
J.~L. Synge, ``The escape of photons from gravitationally intense stars,'' {\em Monthly Notices of the Royal Astronomical Society}, vol.~131, 463(1966).

\bibitem{Luminet1979Image}
J.~P. Luminet, ``Image of a spherical black hole with thin accretion disk,'' {\em Astronomy and Astrophysics}, vol.~75, 228(1979).

\bibitem{bardeen1972rotating}
J.~M. Bardeen, W.~H. Press, and S.~A. Teukolsky, ``Rotating black holes: locally nonrotating frames, energy extraction, and scalar synchrotron radiation,'' {\em Astrophysical Journal}, vol.~178, 370(1972).

\bibitem{chandrasekhar1998themathematical}
S.~Chandrasekhar, {\em The {M}athematical {T}heory of {B}lack {H}oles}.
\newblock Springer Science \& Business Media, (1998).

\bibitem{pahlavon2024effect}
Y.~Pahlavon, F.~Atamurotov, K.~Jusufi, M.~Jamil, and A.~Abdujabbarov, ``Effect of magnetized plasma on shadow and gravitational lensing of a {R}eissner-{N}ordstr{\"o}m black hole,'' {\em Physics of the Dark Universe}, 101543(2024).

\bibitem{hoshimov2024weak}
H.~Hoshimov, O.~Yunusov, F.~Atamurotov, M.~Jamil, and A.~Abdujabbarov, ``Weak gravitational lensing and shadow of a {GUP}-modified {S}chwarzschild black hole in the presence of plasma,'' {\em Physics of the Dark Universe}, vol.~43, 101392(2024).

\bibitem{atamurotov2023quantum}
F.~Atamurotov, M.~Jamil, and K.~Jusufi, ``Quantum effects on the black hole shadow and deflection angle in the presence of plasma,'' {\em Chinese Physics C}, vol.~47, 035106(2023).

\bibitem{atamurotov2021axion}
F.~Atamurotov, K.~Jusufi, M.~Jamil, A.~Abdujabbarov, and M.~Azreg-A{\"\i}nou, ``Axion-plasmon or magnetized plasma effect on an observable shadow and gravitational lensing of a {S}chwarzschild black hole,'' {\em Physical Review D}, vol.~104, 064053(2021).

\bibitem{guo2018observational}
M.~Guo, N.~Obers, and H.~Yan, ``Observational signatures of near-extremal {K}err-like black holes in a modified gravity theory at the event horizon telescope,'' {\em Physical Review D}, vol.~98, 084063(2018).

\bibitem{muhleman1966radio}
D.~Muhleman and I.~Johnston, ``Radio {P}ropagation in the solar gravitational field,'' {\em Physical Review Letters}, vol.~17, 455(1966).

\bibitem{breuer1980propagation}
R.~A. Breuer and J.~Ehlers, ``Propagation of high-frequency electromagnetic waves through a magnetized plasma in curved space-time. {I},'' {\em Proceedings of the Royal Society of London. A. Mathematical and Physical Sciences}, vol.~370, 389(1980).

\bibitem{EventHorizonTelescope:2021srq}
K.~Akiyama {\em et~al.}, ``{First M87 Event Horizon Telescope Results. VIII. Magnetic Field Structure near The Event Horizon},'' {\em Astrophys. J. Lett.}, vol.~910, no.~1, p.~L13, 2021.

\bibitem{akiyama2024first}
K.~Akiyama, A.~Alberdi, W.~Alef, J.~C. Algaba, {\em et~al.}, ``First {S}agittarius {A*} {E}vent {H}orizon {T}elescope {R}esults. {VII}. {P}olarization of the {R}ing,'' {\em The Astrophysical Journal Letters}, vol.~964, 25(2024).

\bibitem{cardoso2019testing}
V.~Cardoso and P.~Pani, ``Testing the nature of dark compact objects: a status report,'' {\em Living Reviews in Relativity}, vol.~22, 4(2019).

\bibitem{muhleman1970radio}
D.~O. Muhleman and R.~D.~o. Ekers, ``Radio interferometric test of the general relativistic light bending near the sun,'' {\em Physical Review Letters}, vol.~24, 1377(1970).

\bibitem{perlick2000ray}
V.~Perlick, {\em Ray Optics, {F}ermat’s principle, and applications to general relativity}.
\newblock Springer Science \& Business Media, (2000).

\bibitem{perlick2017light}
V.~Perlick and O.~Y. Tsupko, ``Light propagation in plasma on {K}err spacetime: separation of the {H}amilton-{J}acobi equation and calculation of the shadow,'' {\em Physical Review D}, vol.~95, 104003(2017).

\bibitem{liu2016electromagnetic}
S.~Liu, ``Electromagnetic fields, size, and copy of a single photon,'' {\em arXiv:1604.03869}, 2016.

\bibitem{houchmandzadeh2020hamilton}
B.~Houchmandzadeh, ``The {H}amilton--{J}acobi equation: {A}n alternative approach,'' {\em American Journal of Physics}, vol.~88, 359(2020).

\bibitem{teo2021spherical}
E.~Teo, ``Spherical orbits around a {K}err black hole,'' {\em General Relativity and Gravitation}, vol.~53, 10(2021).

\bibitem{sheoran2018mass}
P.~Sheoran, A.~Herrera-Aguilar, and U.~Nucamendi, ``Mass and spin of a {K}err black hole in modified gravity and a test of the {K}err black hole hypothesis,'' {\em Physical Review D}, vol.~97, 124049(2018).

\bibitem{kumar2024observational}
S.~Kumar, A.~Uniyal, and S.~Chakrabarti, ``Observational signatures of rotating compact objects in plasma space--time,'' {\em Physics of the Dark Universe}, vol.~44, 101472(2024).

\bibitem{Shapiro1974accretion}
S.~Shapiro, ``Accretion onto black holes: {T}he emergent radiation spectrum. {III}. {R}otating ({K}err) black holes,'' {\em Astrophysical Journal}, vol.~189, 352(1974).

\bibitem{briozzo2023analytical}
G.~Briozzo and E.~Gallo, ``Analytical expressions for pulse profile of neutron stars in plasma environments,'' {\em The European Physical Journal C}, vol.~83, 165(2023).

\bibitem{akiyama2019event}
K.~Akiyama {\em et~al.}, ``Event horizon telescope,'' {\em Astrophys. J. Lett}, vol.~875, L2(2019).

\bibitem{akiyama2022first}
K.~Akiyama, A.~Alberdi, W.~Alef, J.~C. Algaba, R.~Anantua, {\em et~al.}, ``First {S}agittarius {A*} {E}vent {H}orizon {T}elescope results. {I}. {T}he shadow of the supermassive black hole in the center of the {M}ilky {W}ay,'' {\em The Astrophysical Journal Letters}, vol.~930, L12(2022).

\bibitem{johannsen2013systematic}
T.~Johannsen, ``Systematic study of event horizons and pathologies of parametrically deformed {K}err spacetimes,'' {\em Physical Review D}, vol.~87, 124017(2013).

\bibitem{jafarzade2022observational}
K.~Jafarzade, M.~K. Zangeneh, and F.~S. Lobo, ``Observational optical constraints of regular black holes,'' {\em Annals of Physics}, vol.~446, 169126(2022).

\bibitem{hioki2009measurement}
K.~Hioki and K.~Maeda, ``Measurement of the {K}err spin parameter by observation of a compact object’s shadow,'' {\em Physical Review D}, vol.~80, 024042(2009).

\bibitem{wei2013observing}
S.~W. Wei and Y.~X. Liu, ``Observing the shadow of {E}instein-{M}axwell-dilaton-axion black hole,'' {\em Journal of Cosmology and Astroparticle Physics}, vol.~2013, 063(2013).

\bibitem{abdujabbarov2015coordinate}
A.~Abdujabbarov, L.~Rezzolla, and B.~Ahmedov, ``A coordinate-independent characterization of a black hole shadow,'' {\em Monthly Notices of the Royal Astronomical Society}, vol.~454, 2435(2015).

\bibitem{schee2009optical}
J.~Schee and Z.~Stuchl{\'\i}k, ``Optical phenomena in the field of braneworld {K}err black holes,'' {\em International Journal of Modern Physics D}, vol.~18, 1024(2009).

\bibitem{tsukamoto2014constraining}
N.~Tsukamoto, Z.~Li, and C.~Bambi, ``Constraining the spin and the deformation parameters from the black hole shadow,'' {\em Journal of Cosmology and Astroparticle Physics}, vol.~2014, 043(2014).

\bibitem{cunha2015shadows}
P.~V. Cunha, C.~A. Herdeiro, E.~Radu, and H.~F. R{\'u}narsson, ``Shadows of {K}err black holes with scalar hair,'' {\em Physical Review Letters}, vol.~115, 211102(2015).

\bibitem{younsi2016new}
Z.~Younsi, A.~Zhidenko, L.~Rezzolla, R.~Konoplya, and Y.~Mizuno, ``New method for shadow calculations: {A}pplication to parametrized axisymmetric black holes,'' {\em Physical Review D}, vol.~94, 084025(2016).

\bibitem{wang2018chaotic}
M.~Wang, S.~Chen, and J.~Jing, ``Chaotic shadow of a non-{K}err rotating compact object with quadrupole mass moment,'' {\em Physical Review D}, vol.~98, 104040(2018).

\bibitem{gebhardt2011black}
K.~Gebhardt, J.~Adams, D.~Richstone, T.~R. Lauer, S.~M. Faber, K.~G{\"u}ltekin, J.~Murphy, and S.~Tremaine, ``The black hole mass in {M87} from {G}emini/{NIFS} adaptive optics observations,'' {\em The Astrophysical Journal}, vol.~729, 119(2011).

\bibitem{walsh2013the}
J.~L. Walsh, A.~J. Barth, L.~C. Ho, and M.~Sarzi, ``The {M87} black hole mass from gas-dynamical models of space telescope imaging spectrograph observations,'' {\em The Astrophysical Journal}, vol.~770, 86(2013).

\bibitem{johannsen2011metric}
T.~Johannsen and D.~Psaltis, ``Metric for rapidly spinning black holes suitable for strong-field tests of the no-hair theorem,'' {\em Physical Review D}, vol.~83, 124015(2011).

\bibitem{kumar2020black}
R.~Kumar and S.~G. Ghosh, ``Black hole parameter estimation from its shadow,'' {\em The Astrophysical Journal}, vol.~892, 78(2020).

\bibitem{takahashi2004shapes}
R.~Takahashi, ``Shapes and positions of black hole shadows in accretion disks and spin parameters of black holes,'' {\em The Astrophysical Journal}, vol.~611, 996(2004).

\bibitem{grenzebach2014photon}
A.~Grenzebach, V.~Perlick, and C.~L{\"a}mmerzahl, ``Photon regions and shadows of {K}err-{N}ewman-{N}{U}{T} black holes with a cosmological constant,'' {\em Physical Review D}, vol.~89, 124004(2014).

\bibitem{jafarzade2024kerr}
K.~Jafarzade, S.~H. Hendi, M.~Jamil, and S.~Bahamonde, ``Kerr--{N}ewman black holes in {W}eyl--{C}artan theory: {S}hadows and {EHT} constraints,'' {\em Physics of the Dark Universe}, vol.~45, 101497(2024).

\bibitem{jafarzade2024thermodynamics}
K.~Jafarzade, B.~E. Panah, and M.~Rodrigues, ``Thermodynamics and optical properties of phantom {AdS} black holes in massive gravity,'' {\em Classical and Quantum Gravity}, vol.~41, 065007(2024).

\bibitem{wang2022shadow}
H.~Wang and S.~Wei, ``Shadow cast by {K}err-like black hole in the presence of plasma in {E}instein-bumblebee gravity,'' {\em The European Physical Journal Plus}, vol.~137, 17(2022).

\end{thebibliography}
\end{document}